\documentclass[%
 reprint,
superscriptaddress,
 amsmath,amssymb,
 aps,
pre,
]{revtex4-2}

\usepackage{graphicx}% Include figure files
\usepackage{dcolumn}% Align table columns on decimal point
\usepackage{bm}% bold math
\usepackage{bbm}
\usepackage{hyperref}
\usepackage{color}

\newcommand{\figref}[2]{\hyperref[#1]{\ref*{#1}#2}}

\begin{document}

\preprint{APS/123-QED}

\title{A Viscoelastic Theory for Ultrasound-Induced Intracellular Streaming}

\author{Niels Gieseler}
\affiliation{Institute for Theoretical Physics, Heidelberg University, Philosophenweg 19, 69120 Heidelberg, Germany}
\affiliation{
BioQuant, Heidelberg University, Im Neuenheimer Feld 267, 69120 Heidelberg, Germany}
\affiliation{Max Planck Institute for Medical Research, Jahnstrasse 29, 69120 Heidelberg, Germany
}
\author{Falko Ziebert}
\affiliation{Institute for Theoretical Physics, Heidelberg University, Philosophenweg 19, 69120 Heidelberg, Germany}
\affiliation{
BioQuant, Heidelberg University, Im Neuenheimer Feld 267, 69120 Heidelberg, Germany}
 \author{Ulrich S. Schwarz}
 \email{schwarz@thphys.uni-heidelberg.de}
\affiliation{Institute for Theoretical Physics, Heidelberg University, Philosophenweg 19, 69120 Heidelberg, Germany}
\affiliation{
BioQuant, Heidelberg University, Im Neuenheimer Feld 267, 69120 Heidelberg, Germany}%

\date{\today}% It is always \today, today,
             %  but any date may be explicitly specified

\begin{abstract}
Ultrasound is increasingly used to control biological cells, but its multiple physical effects, including radiation force, streaming, heating, and cavitation, require theoretical frameworks that can quantify their relative strengths.
Here we develop a semi-analytical model 
to predict flow patterns and energy distributions
of ultrasound-induced intracellular streaming.
Cells are modeled as viscoelastic droplets immersed
in a fluid with different viscoelastic properties. 
For this geometry, the momentum equation with an Oldroyd-B
constitutive law is solved using a 
perturbation expansion commonly applied in
acoustofluidics. The resulting equations for the different orders of the expansion are solved using partial wave expansion and explicit integration. We investigate our model for a range of parameters which is relevant both for synthetic
polymer and for protein solutions. 
We find a series of flow reversals that we can connect to the distribution of energy into the different modes.
We also consider density and compressibility contrasts, which shift these transitions and introduce
new ones. Finally, we discuss the potential 
biological relevance of our mathematical results 
for cellular sensing and signaling. 
\end{abstract}

%Use showkeys class option if keyword
                              %display desired
\maketitle

\section{Introduction}

Ultrasound is not only a very important imaging modality
in biomedicine \cite{10.1093/oso/9780195168310.001.0001}, but also a powerful tool for manipulating biological cells \cite{qu2026sono}.
Its main physical effects used in these applications include acoustic radiation force, acoustic streaming, heating and cavitation, whose prediction is challenging due to the strongly
non-linear nature of the underlying equations \cite{Hamilton2024-hw, HUMPHREY2007195}.
Recently, significant progress has been made in exploiting ultrasound-induced microscale forces and hydrodynamic flows for applications ranging from label-free particle manipulation and trapping \cite{C6LC00502K,Baudoin2020,mi14081487,VanAssche2020} to the biophysical control of cells and soft matter systems \cite{Athanassiadis2022, Melde2024, Huang2024}. 

The potential applications of ultrasound
on cells are as varied as are the physical effects connected to ultrasound. 
One potential application of ultrasound is in the field of drug delivery, where the goal is to control the release of a drug and its uptake by the targeted cell in both space and time \cite{Athanassiadis2022}. Microbubbles have been used as the most common drug carrier that can be ruptured with high spatial control in tissue using focused ultrasound \cite{Stride2019, D1MA01197A}. Recently, antibubbles have been proposed to increase the amount of drug per carrier compared to bubbles \cite{Nico2023}. Furthermore, ultrasound-induced  mechanochemistry can be used to control not only drug release, but also drug activation \cite{D2SC05196F}. Another
interesting option is sonoporation, the permeabilization of the cell membrane using a targeted ultrasound stimulus, thereby increasing drug uptake. Traditional sonoporation methods typically require microbubble ultrasound contrast agents and rely mostly on cavitation effects \cite{doi:10.34133/2022/9807347, KOOIMAN20201296, Marmottant2003}. However, due to the risk of cell and cargo damage, other methods have been investigated \cite{Rich2022}. In both approaches, it is not fully clear whether acoustic microstreaming, radiation forces, jetting, or other effects are the dominant mechanical effect. Theoretical analysis is further complicated by the influence of the complicated soft matter system on bubble dynamics \cite{annurev:/content/journals/10.1146/annurev-fluid-010518-040352}.

A second active field of research is ultrasound cell stimulation. Due to the ability of ultrasound to be focused in deep tissue, this opens several therapeutic avenues such as enhanced bone healing, immunomodulation or neurostimulation \cite{bonehealing, RIX2024146, PELLOW2024734}. A large amount of work has been focused on neurostimulation. Due to the ability of ultrasound to be focused transcranially in deep brain regions using holograms \cite{Burstow_2025}, it is a prime contender for noninvasive neurostimulation \cite{Meng2021}. Recent studies have shown various effects in animals and humans \cite{PELLOW2024734, 10.3389/fnhum.2021.749162}, such as modulation of pain perception \cite{BADRAN20201805}, mood \cite{HAMEROFF2013409}, and improved neuropsychological scores in patients with Alzheimer's disease \cite{Beisteiner2020}.
However, regardless of the application, the underlying mechanical pathways to cell stimulation are not fully understood. The interaction of ultrasound with tissue and single cells generates multiple co-occurring effects, such as forces generated by the microstreaming, acoustic radiation forces, or cavitation. There are several theories to explain how ultrasound affects cells, such as the deflection of the cell membrane or activation of mechanosensitive ion channels \cite{10.3389/fphy.2020.00150, BLACKMORE20191509}. Considerable interest is centered on identifying ion channels that mediate the cell response \cite{Yoo2022, Ye2018}.

A third and strongly growing application area for ultrasound is sonogenetics \cite{10.3389/fphys.2020.00787, 10.3389/facou.2023.1269867, 11249725, 10.1002/anie.202317112}. 
By using cells that have been genetically modified, sonogenetics combines the noninvasive spatial control of ultrasound and the high specificity of genetics. Sonogenetics was first introduced as the engineered expression of mechanosensitive ion channels to induce or increase the sensitivity to ultrasound for specific cells \cite{Ibsen2015}. However, to further progress the capabilities of the method, all types of mechanotransduction pathways have to be investigated in the diverse context of cellular mechanobiology \cite{qu2026sono}. In addition to engineering the sensitivity of specific cells, sonogenetics also offers the possibility to steer gene expression using downstream transcription \cite{10.3389/facou.2023.1269867, 10.1002/anie.202317112}. While the main focus of the field is on mechanical pathways, it is also possible to steer gene expression using ultrasound heating, by coupling the heat shock protein to, for example, CAR-T cell signaling \cite{Liu2025}. 

A less explored mechanical pathway is the control of intracellular streaming. 
Active streaming driven by molecular motors occurs in many biological systems and often is associated with the establishment of body polarity \cite{goldstein2015physical}. Using focused-light-induced cytoplasmic streaming (FLUCS), it has been shown that such control can be also achieved from the outside, 
e.g. to revert polarity in \textit{C. elegans} embryos \cite{FLUCS}. Disruption of
biological streaming can have dramatic consequences in development, such as \textit{situs inversus},
where the heart ends up on the wrong side of the body \cite{nonaka2005novo}.
Controlling intracellular streaming by ultrasound, therefore, opens the door
to control important biological processes.

Historically, acoustic intracellular streaming has been studied mainly for plant cells \cite{HARVEY1928,5008040,Gershoy1976,MARTIN1978131,Martin_1979}, 
possibly because they are easier to manipulate than animal cells. 
Stable cavitation from gas bubbles in plant cells was shown to be important in generating flows \cite{Martin_1979, Gershoy1976}, the absence of which makes the direct comparison to animal cells challenging \cite{MARTIN1978131}. Early attempts to create acoustic streaming flow in live animal cells concluded that, due to the high internal cohesion of animal cells, cells rupture before large scale internal streaming occurs \cite{10.1121/1.1913158}. A recent experimental study, however, showed that cells can produce an extracellular microstreaming flow, and reported that in cells exposed to a small hypotonic shock, streaming flows can extend into the inside of the cells \cite{Dancing_with_the_Cells}. Additionally, acoustic streaming in giant unilamellar vesicles (GUVs) was investigated both experimentally and using simulations \cite{Pereno2020}. However, the model considered a purely viscous model and only the streaming generated by the resonator mode in the bulk medium, not the microstreaming generated by the geometry of the cell itself. 

In order to mathematically model acoustic streaming in and around biological
cells, one has to carefully consider the relevant length and time scales.
Finding a general model for all applications is challenging due to the complexity of a cell. When discussing cellular rheology, there are two different viewpoints depending on the length scale.
On the one hand, one can aim at modeling the cell as one object, which requires
a decision to which mechanical
model is most appropriate in regard to whole cell mechanics
\cite{mietke2015extracting}. Within recent years, there has been a large development in microfluidic cytometry methods, such as constriction-based methods \cite{Otto2015, Fregin2019} or shear flow cytometry \cite{Gerrum}, which have a throughput advantage over more established methods such as AFM-indentation or micropipette aspiration. The approximation of a cell as homogeneous, which is usually applied in these methods, is typically well justified \cite{Wohlrab2024}. Nevertheless, there are large differences in the models applied to describe homogeneous cells, from elastic \cite{Otto2015}, Kelvin-Voigt \cite{Fregin2019} to power-law rheology \cite{Gerrum}.

On the other hand, when considering intracellular streaming, the relevant property becomes the local rheology of the cytoplasm. Experimentally, this is challenging to measure, as the cytoplasm is shielded by the cortex. Common methods are particle tracking microrheology \cite{TSENG20023162,PARRY2014183,Smigiel2022,VALENTINE2005680}, pulling of magnetically \cite{oldroyd} or optically \cite{doi:10.1073/pnas.1702488114, Hurst2021} trapped beads, or magnetic rotation \cite{Berret2016}. The models applied in these measurements vary depending on the length and time scale, and span from purely viscous media \cite{Smigiel2022}, to viscoelastic media \cite{Hurst2021,oldroyd,VALENTINE2005680,TSENG20023162,Berret2016}, poroelastic networks \cite{Moeendarbary2013, doi:10.1073/pnas.1702488114}, to soft glassy media \cite{PARRY2014183}. Of these, viscoelastic media are a good approximation for acoustofluidic simulations, as they offer a more accurate description than the purely viscous models often used in numerical cell models \cite{FLUCS,10.1371/journal.pcbi.1006588,Pereno2020}, while avoiding the complicated models associated with porous environments. Several viscoelastic models have been considered for the description of the cytoplasm, including the generalized Maxwell model \cite{Berret2016}, the fractional Kelvin-Voigt model \cite{Hurst2021}, Zener models \cite{OVALLEFLORES2023105734}, or Oldroyd-B fluids \cite{oldroyd}. 

Here, we advance a theoretical framework for understanding the subcellular effects of ultrasound sonication, which could eventually be leveraged to control subcellular processes within biological cells. To this end, we develop semi-analytical solutions for acoustic microstreaming inside a viscoelastic spherical particle or droplet, immersed in a viscoelastic fluid, disregarding
the effect of the nucleus and other organelles. The calculation of intracellular streaming thus closely follows the classical acoustofluidic problem of microstreaming flows surrounding a spherical particle or bubble. Initial experimental observations of these flows surrounding bubbles were done by Elder \cite{10.1121/1.1907611}. Riley \cite{10.1093/qjmam/19.4.461} was among the first to calculate the streaming flows around a translationally oscillating rigid sphere. In a later work, Davidson and Riley \cite{DAVIDSON1971217} extended this theory to bubbles by introducing a slip condition on the surface boundary. The effect of pulsation, crucial to bubble dynamics, was later treated by Wu and Du \cite{10.1121/1.418223} as well as Longuet-Higgins \cite{10.1098/rspa.1998.0183}.
While there is a large body of work on the streaming flows outside of the particle, only few studies have investigated the streaming inside the bubble or fluid particle \cite{10.1121/1.418223, 10.1121/1.428182, 10.1093/qjmam/hbl007, Thierry_SiB,10.1063/5.0315990}. Baasch \textit{et al.} \cite{Thierry_SiB} first introduced a solution for the streaming flows inside and outside of a fluid particle undergoing monopole and dipole oscillations, which did not impose any restriction on the ratio of the viscous boundary layer to the particle radius, by introducing a partial wave expansion. Recently, this work was revisited by Doinikov \textit{et al.} \cite{10.1063/5.0315990} and generalized to all axissymetric multipolar modes. Additionally, Doinikov \textit{et al.} \cite{VS_Doinikov} applied the same approach to calculate the streaming in a viscoelastic Oldroyd-B fluid surrounding an elastic particle, undergoing oscillations of all multipolar modes. Numerically, the streaming around nonspherical particles can be calculated using either finite element simulations for first- and second-order flows \cite{PhysRevE.106.015105}, or direct flow simulations for the entire field \cite{doi:10.1126/sciadv.adh5260}. 

As the size of a physiological cell can be comparable to the viscous boundary layer thickness at the MHz-scale, we closely follow the work by Baasch \textit{et al.} \cite{Thierry_SiB} which imposes no limits on their ratio. Here we extend
the initial work by Doinikov \textit{et al.} \cite{VS_Doinikov} on Oldroyd-B fluids to include viscoelastic effects 
both inside and outside, and apply the resulting theory in the context of biophysics and internal streaming in biological cells. Furthermore, we study the distributions of the kinetic energy of the streaming flows and show that it can be used to investigate in detail the occurring flow transitions and reversals.

\section{Theory}

\subsection{Model equations}

We first introduce our model and its central equations. 
The solution is split into three parts. First, we perform a multiscale series expansion, as detailed in Sect.~\ref{sec::series}, which splits the problem into a first order that describes the acoustic waves and a second order that describes the streaming. The acoustic waves are treated using scattering theory in Sect.~\ref{sec::first_order}, whereas the streaming requires the solution of two differential equations with inhomogeneities involving the first order, as detailed in Sect.~\ref{sec::second_order}.

\begin{figure}[t!]
     \centering
 	\includegraphics[width=0.99\linewidth, trim={5cm 14.9cm 5cm 4.5cm},clip,angle=0]{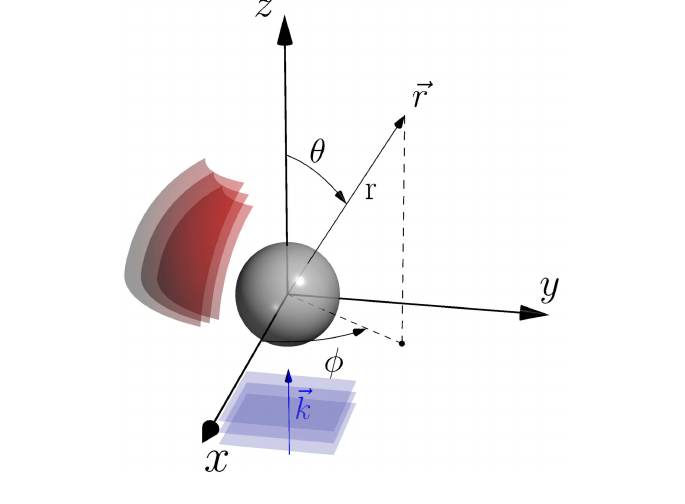}
 	    \caption{Representation and geometry of the system. A spherical domain of radius $R$, shown in grey, is placed in a surrounding medium and sonicated by a plane wave aligned with the $z$-axis. The incoming plane wave and the scattered spherical wave are shown in blue and red, respectively. Both media, the inside of the particle and the outside, are described by viscoelastic Oldroyd-B models with different parameters.\label{img::representation}} 
\end{figure}

In the following, we consider a spherical particle or droplet placed in an infinite medium and irradiated with a plane acoustic wave. Both sphere and surrounding medium are characterized by viscoelastic Oldroyd-B constitutive laws, however with different material parameters. 
The aim is to calculate the resulting acoustic streaming flows (which we will term just streaming in the following), both inside and outside the particle. Note that this includes both the bulk streaming and the boundary-driven microstreaming. Due to the symmetry, we will use spherical polar coordinates $(r,\theta,\phi)$ centered on the particle. Without restriction, we orient the incoming wave along the $z$-axis, resulting in 
an axial symmetry around $\phi$. The setup is shown in Fig.~\ref{img::representation}.

The behavior of the fluids is described using the balance equations for mass and momentum
\begin{align}
    \partial_t\rho&=-\nabla\cdot(\rho\vec{v}),\label{eq::mass_continuity_general}\\
    \rho(\partial_t\vec{v}+\vec{v}\cdot\nabla\vec{v})&=\nabla \cdot\sigma.\label{eq::momentum_continuity_general}
\end{align}
Here $\rho$ and $\vec{v}$ are the fluid density and velocity, respectively, and 
$\sigma$ is the stress tensor. In the following, we use the convention $(\nabla\vec{v})_{ij}=\partial_iv_j$ for the velocity gradient tensor and $(\nabla\cdot\sigma)_i=\partial_j\sigma_{ij}$ for the tensor divergence. Note that density, velocity, and stress tensor are all functions of both space and time. 
However, except when explicitly stated, the dependencies are dropped in the notation for sake of brevity. 

For a Newtonian fluid, the momentum equation becomes the Navier-Stokes equation,
because then the stress tensor $\sigma$ has only three
simple contributions, involving pressure,  shear and
bulk viscosity. 
In this work, we consider the Oldroyd-B model, which can be described 
as a sum of a Newtonian viscous fluid 
and an additional viscoelastic contribution
\begin{align}
    \sigma=&-p\mathbbm{1}+\eta_f\left[\nabla\vec{v}+(\nabla\vec{v})^T
    \hspace{-1mm}-\frac{2}{3}(\nabla\cdot\vec{v})\mathbbm{1}\right]\nonumber\\
    &+\xi_f(\nabla\cdot\vec{v})\mathbbm{1}+\mathbf{\mathbf{\tau}}.\label{eq::sigma_general}
\end{align}
Here $p$ is the pressure field and  
$\eta_f$ and $\xi_f$ are the shear and bulk viscosities of the fluid, respectively. 
$\mathbf{\tau}$ is an additional viscoelastic stress tensor of second rank, which we will call polymer stress in the following, as usual in the
Oldroyd-B model literature. 
$\mathbbm{1}$ is the identity tensor, and a superscript $^T$ denotes the transpose of a tensor. 

In the Oldroyd-B model, the polymer stress is given by an upperconvected Maxwell fluid model
\begin{align}
    &\mathbf{\mathbf{\tau}}+\lambda_M\left[\frac{\partial \mathbf{\mathbf{\tau}}}{\partial t}+\vec{v}\cdot\nabla\mathbf{\mathbf{\tau}}-\mathbf{\mathbf{\tau}}\cdot\nabla\vec{v}-(\nabla\vec{v})^T\cdot\mathbf{\mathbf{\tau}}\right]\nonumber\\
&=\eta_p\left[\nabla\vec{v}+(\nabla\vec{v})^T-\frac{2}{3}(\nabla\cdot\vec{v})\mathbbm{1}\right]+\xi_p(\nabla\cdot\vec{v})\mathbbm{1}.
    \label{eq::tau_general}
\end{align}
Note that this is a dynamic equation for the polymer stress $\tau$. The shear and bulk viscosities of the polymer contribution are denoted with $\eta_p$ and $\xi_p$, respectively. The Maxwell time $\lambda_M$ is the characteristic time scale describing relaxation. The convected derivative in the second term on the left-hand side is needed to ensure frame invariance and depends sensitively on the convention of the velocity gradient. 

Schematically, the Oldroyd-B model can be thought of as a single dashpot in parallel to a combination of a spring and a second dashpot in series. This results in three characteristic regimes: For the shortest timescales, the spring buffers the second dashpot, and the behavior is viscous with 
viscosity $\eta_f$. For very long time scales, the spring always equilibrates, and the behavior is again viscous, but now 
with the total viscosity $\eta=\eta_f+\eta_p$. Between these two limiting cases, the behavior is predominantly elastic. 
Hence, the model has two characteristic time scales: the Maxwell time $\lambda_M$ sets the transition from the elastic to the long time scale viscous regime, and the 
retardation time 
$\lambda_r=\lambda_M\frac{\eta_f}{\eta_f+\eta_p}$ determines the transition between the elastic and the short time scale viscous regime.

In the following sections, we solve Eqs.~(\ref{eq::mass_continuity_general})
-(\ref{eq::tau_general}), for the situation
of a spherical particle sonicated by a plane wave as
sketched in Fig.~\ref{img::representation},
using a series expansion.

\subsection{Series Expansion\label{sec::series}}

To solve the set of nonlinear equations of our model, we use the perturbation expansion approach common to acoustofluidics \cite{bruus2, Nyborg2024}. We set 
\begin{align}
    \rho&=\rho^{(0)}+\rho^{(1)}+\rho^{(2)},\nonumber\\
    p&=p^{(0)}+c^2\rho^{(1)}+p^{(2)},\nonumber\\
    \vec{v}&=\vec{v}^{(1)}+\vec{v}^{(2)}\label{eq::expansion},
\end{align}
where we already applied the adiabatic approximation $p^{(1)}=c^2\rho^{(1)}$, and assumed a quiescent fluid implying $\vec{v}^{(0)}=0$. With this slow streaming approximation, the acoustic wave and the resulting streaming are separated in the first and second orders of the velocity, respectively. 
However, with large acoustic intensities, the streaming does not necessarily have to be small with respect to the acoustic velocity \cite{friend}. Therefore, any solution calculated using the slow streaming approximation should be tested for consistency by comparing the resulting velocities of both orders to ensure that the streaming is indeed small.

\subsection{First Order Solution: Acoustic Waves\label{sec::first_order}}

Following the series expansion from the last section, we now solve the first-order equations. To this end, as a first step, we derive the governing equations for the first-order velocity field, which will result in acoustic waves. In a second step, we will then solve these equations using a partial wave expansion.

\subsubsection{Governing Equations}
The key assumption,
based on the expectation that one recovers acoustic waves from the first-order expansion, 
is that all first-order fields oscillate 
harmonically in time like 
$e^{-i\omega t}$.
%, which simplifies the time derivatives. 
Inserting the series expansion 
Eq.~(\ref{eq::expansion}) in Eqs.~(\ref{eq::mass_continuity_general})-(\ref{eq::tau_general}) and collecting first order terms, we arrive at 
\begin{align}
    -i\omega\rho^{(0)}\vec{v}^{(1)}=&\frac{i\rho^{(0)}c^2}{\omega}\nabla \left(\nabla\cdot\vec{v}^{(1)}\right)+\eta_c\nabla^2\vec{v}^{(1)}\nonumber\\
    &+\left(\xi_c+\frac{\eta_c}{3}\right)\nabla(\nabla\cdot\vec{v}^{(1)}).\label{eq::first_order_velocity_dgl}
\end{align}
Here, we defined the complex viscosities as
\begin{align}
    \eta_c=&\eta_f+\frac{\eta_p}{1-i\omega\lambda_M},\label{eq::eta_c}\\
    \xi_c=&\xi_f+\frac{\xi_p}{1-i\omega\lambda_M.}\label{eq::xi_c}.
\end{align}
Note that the viscous response is governed by the Deborah number $\mathrm{De}=\omega\lambda_M$ in the denominator, which shows the relation of the relaxation to the probing time.

We continue by introducing a Helmholtz decomposition for the velocity
\begin{align}
    \vec{v}^{(1)}=\nabla\varphi^{(1)}+\nabla\times\vec{\psi}^{(1)},\label{eq::def_potentiale_first_order}
\end{align}
where the scalar potential $\varphi^{(1)}$ represents the compressional mode of the velocity, dominant in the long range, and the vector potential $\vec{\psi}^{(1)}$ represents the shear flow necessary to fulfill the boundary conditions on the surface of the particle.
Note that because of the axial symmetry,
one has $\varphi^{(1)}(\vec{r})=\varphi^{(1)}(r,\theta)$ and $\vec{\psi}^{(1)}(\vec{r})=\psi^{(1)}(r,\theta)\hat{e}_\phi$.

By either applying a divergence or a rotation on Eq.~(\ref{eq::first_order_velocity_dgl}), the two potentials can be decoupled into two separate Helmholtz equations
\begin{align}
    \nabla^2\varphi^{(1)}+k_f^2\varphi^{(1)}&=0,\label{eq::first_order_helmholtz_phi}\\
    \nabla^2\vec{\psi}^{(1)}+k_\nu^2\vec{\psi}^{(1)}&=0,\label{eq::first_order_helmholtz_psi}
\end{align}
with the compressional and shear wavenumber
\begin{align}
    k_f&=\frac{\omega}{c}\left[1-\frac{i\omega}{\rho^{(0)}c^2}\left(\xi_c+\frac{4\eta_c}{3}\right)\right]^{-\frac{1}{2}},\label{eq::kf}\\
    k_\nu&=(1+i)\sqrt{\frac{\rho^{(0)}\omega}{2\eta_c}}.\label{eq::knu}
\end{align}
Thus, the first-order equations reduce to two separate wave equations for both shear and compressional waves, which validates the time-harmonic assumption. Note that these are functionally the same differential equations as for a purely viscous medium in \cite{Thierry_SiB}, with the shear and bulk viscosity replaced by the complex viscosities Eqs.~(\ref{eq::eta_c}) and (\ref{eq::xi_c}).

For an ultrasound source with a frequency of $1\,\mathrm{MHz}$, a typical order of magnitude for neurostimulation applications \cite{10.3389/fnhum.2021.749162, BLACKMORE20191509}, the longitudinal/compressional
wavelength $\lambda_f=\frac{2\pi}{\Re(k_f)}$ in water is $1.5\,\mathrm{mm}$ and thus much larger than the typical size of a cell at a diameter of $30\,\mathrm{\mu m}$. 
The shear wave length, $\lambda_\nu=\frac{2\pi}{\Re(k_\nu)}$, however, is only $3.5\,\mathrm{\mu m}$ and thus comparable to the size of a cell. Note that while for a viscous medium the shear wavelength and 
the viscous boundary layer 
thickness $\delta$
are connected by $\lambda_\nu=2\pi\delta$, this is not the case in the present viscoelastic case due to the complex viscosities. 

In the following, we solve the first order problem using a partial wave expansion. 
We then turn to the second-order equations, where the first-order solutions will 
act as the inhomogeneity.

\subsubsection{Solution using Scattering Theory}

As the first-order equations reduce to two wave equations, the resulting problem 
for the given geometry 
is the scattering of a wave from a 
spherical particle.  
A typical treatment is the partial wave expansion: we expand the incoming field, 
the field inside the spherical domain, 
and the scattered field in Legendre polynomials and spherical Bessel functions, the eigenfunctions of the Helmholtz equation, and calculate the expansion coefficients from the boundary conditions. 

In this work, we consider an incident 
compressional plane wave, comprised only by the scalar potential 
$\vec{v}_{\mathrm{ac}}=\nabla\phi_\mathrm{ac}$ that can be expanded into eigenfunctions as
\begin{align}
    \varphi_\mathrm{ac}^{(1)}=e^{-i\omega t}\sum_{n=0}^\infty A_n j_n(k_fr)P_n(\mu),
\end{align}
due to the axial symmetry. Here $j_n$ are the spherical Bessel functions, $P_n$ are the Legendre polynomials, and $\mu=\cos\theta$. In this work, we focus on a standing wave, for which the expansion coefficients $A_n$ are given by $A_n=\frac{1}{2}A(2n+1)i^n\left[e^{ik_fd}+(-1)^ne^{-ik_fd}\right]$, where $A$ is the amplitude of the velocity potential and $d$ is the distance of the sphere to the velocity node of the incoming wave. The framework, however, can also be applied to plane propagating waves, in which case $A_n=A(2n+1)i^n$. The relation between the amplitude of the velocity potential and the pressure of the incoming wave is given by
\begin{align}
    A=\frac{1-\tfrac{i\omega}{\rho^{(0)}c^2}\left(\xi_c+\tfrac{4\eta_c}{3}\right)}{i\omega\rho^{(0)}}p_\mathrm{ac},
\end{align}
including the corrections due to the viscosity of the medium.

In contrast to the incoming field, the scattered field (outside the spherical domain)
does have shear contributions, and its expansions are
\begin{align}
    \varphi_\mathrm{sc}^{(1)}&=e^{-i\omega t}\sum_{n=0}^\infty a_n h_n^{(1)}(k_fr)P_n(\mu),\\
    \psi_\mathrm{sc}^{(1)}&=e^{-i\omega t}\sum_{n=1}^\infty b_n h_n^{(1)}(k_\nu r)P_n^1(\mu),\label{eq::expansion_b_n}
\end{align}
where $h_n^{(1)}$ are the spherical Hankel functions of the first kind, $P_n^1$ are the associated Legendre polynomials with $m=1$, and $a_n$ and $b_n$ are the scattering coefficients. The Hankel functions are needed to fulfill the Sommerfeld radiation condition of radiated energy only propagating outwards \cite{SOMMERFELD1949166}.
Note that the scalar magnitude of the vector potential has to be expanded in associated Legendre polynomials, as the vector potential follows the vectorial Helmholtz equation, not the scalar one.

Finally, the field inside the spherical domain also has to be expanded as
\begin{align}
    \varphi_\mathrm{in}^{(1)}&=e^{-i\omega t}\sum_{n=0}^\infty \hat{a}_n j_n(\hat{k}_fr)P_n(\mu),\\
    \psi_\mathrm{in}^{(1)}&=e^{-i\omega t}\sum_{n=1}^\infty \hat{b}_n j_n(\hat{k}_\nu r)P_n^1(\mu),\label{eq::expansion_b_n_hat}
\end{align}
where the spherical Bessel functions are used for their regularity at 0. In the following, we denote all material parameters and scattering coefficients referring to quantities inside the sphere 
with a hat (" $\hat{}$ ").

By inserting the expanded potentials in the Helmholtz decomposition, Eq.~(\ref{eq::def_potentiale_first_order}), it can be shown that the radial and tangential first-order velocity also decompose into eigenmodes as
\begin{align}
    v_r^{(1)}(r,\theta,t)=e^{-i\omega t}\sum_{n=0}^{\infty}v^{(1)}_{rn}(r)P_n(\mu),\label{eq::app::vr1multipole}\\
    v_\theta^{(1)}(r,\theta,t)=e^{-i\omega t}\sum_{n=1}^{\infty}v^{(1)}_{\theta n}(r)P_n^1(\mu),\label{eq::app::vtheta1multipole}
\end{align}
where the radial functions are given by
\begin{align}
    v^{(1)}_{rn}(r)&=\frac{\partial\varphi^{(1)}_n(r)}{\partial r}-\frac{n(n+1)}{r}\psi^{(1)}_n(r),\label{eq::V1r}\\
    v^{(1)}_{\theta n}(r)&=\frac{\varphi^{(1)}_n(r)-\psi^{(1)}_n(r)}{r}-\frac{\partial\psi^{(1)}_n(r)}{\partial r}.\label{eq::V1theta}
\end{align}

To fully determine the fields, the scattering coefficients $a_n, b_n, \hat{a}_n, \mathrm{and}~ \hat{b}_n$ have to be determined. This is done by imposing boundary conditions on the surface of the particle, namely
continuity of both velocity and stress in both the radial and tangential directions. 
Due to the orthogonality of the Legendre polynomials, the boundary conditions reduce to a linear system of equations for each order $n$. The details of the calculations
are shown in Appendix~\ref{App::scattering}. 

In the end, one obtains an analytic solution for the first-order velocity as an infinite series. However, the strength of the partial wave expansion is that the series converges quickly if $\Re(k_f)R<1$,
where 
$k_f$ is the compressional wave number and
$R$ the radius of the domain,
and that it hence can be truncated at a suitable order $N_\mathrm{max}$. 

%Having finished the calculation of the first-order velocity, we will now treat the second-order expressions where the solutions we just derived act as inhomogeneities.

\subsection{Second Order Solution: Acoustic Streaming\label{sec::second_order}}

Acoustic streaming is a non-vanishing time-averaged flow that arises when an acoustic wave propagates through a medium. This flow is a consequence of the nonlinearity of the momentum equation, Eq.~(\ref{eq::momentum_continuity_general}), present in the second-order approximation. Because we are explicitly only interested in the acoustic streaming, the second-order equations are solved in their time-averaged form. Additionally, here we investigate the case where the streaming flow is steady, so transients are not accounted for. As a consequence of assuming that the time-averaged flow is steady and thus does not depend on time, the perturbation series results in the use of multiple time scales between the different expansion orders. While the relevant time scale of the first-order equations is dictated by the frequency of the incoming wave, the second-order equations are solved in the long time scale limit of the Oldroyd-B model.   

Therefore, we calculate the expansion of Eqs.~(\ref{eq::mass_continuity_general})-(\ref{eq::tau_general}) up to second order, and apply a time average, assuming the second-order expressions to be constant in time after averaging. 
After simplification, this leads to 
\begin{align}
    \nabla\cdot\langle\vec{v}^{(2)}\rangle=-\frac{1}{\rho^{(0)}}\nabla\cdot\langle\rho^{(1)}\vec{v}^{(1)}\rangle,\label{eq::second_order_contiuity}\\
    (\eta_f+\eta_p)\nabla^2\langle\vec{v}^{(2)}\rangle+\left(\xi_f+\xi_p+\frac{\eta_f+\eta_p}{3}\right)\nabla(\nabla\cdot\langle\vec{v}^{(2)}\rangle)\nonumber\\-\nabla\langle p^{(2)}\rangle=\rho^{(0)}\langle\vec{v}^{(1)}\nabla\cdot\vec{v}^{(1)}+\vec{v}^{(1)}\cdot\nabla\vec{v}^{(1)}\rangle+\lambda_M\langle\nabla\cdot\mathbf{T}\rangle,\label{eq::second_order_momentum_stress_inserted}
\end{align}
for the second-order balances of mass and momentum,
respectively. 
Here $\langle...\rangle$ denotes the time average. In the expression of the second-order momentum, we introduced the tensor
$\mathbf{T}$ as 
\begin{align}
    \mathbf{T}=\vec{v}^{(1)}\cdot\nabla\mathbf{\tau}^{(1)}-\mathbf{\tau}^{(1)}\cdot\nabla\vec{v}^{(1)}-(\nabla\vec{v}^{(1)})^T\cdot\mathbf{\tau}^{(1)}.\label{eq::T}
\end{align}
It describes the second-order contribution of the convected derivative terms 
in Eq.~(\ref{eq::tau_general}) and involves
the first-order polymer stress, 
\begin{align}
    \mathbf{\mathbf{\tau}}^{(1)}=& \,\frac{\eta_p}{1-i\omega\lambda_M}\left[\nabla\vec{v}^{(1)}+(\nabla\vec{v}^{(1)})^T-\frac{2}{3}(\nabla\cdot\vec{v}^{(1)})\mathbbm{1}\right]\nonumber\\
    &+\frac{\xi_p}{1-i\omega\lambda_M}(\nabla\cdot\vec{v}^{(1)})\mathbbm{1},
\label{eq::first_order_polymer_stress_prelim}
\end{align}
that is calculated from 
Eq.~(\ref{eq::tau_general}) 
using the time-harmonic assumption. 

In Eq.~(\ref{eq::second_order_momentum_stress_inserted}), the viscosity  is the sum of fluid and polymer contributions, 
since the steady time average
in the second order relates to the long time scale limit of the Oldroyd-B model. Note that in contrast, the complex viscosities, Eqs~(\ref{eq::eta_c}) and (\ref{eq::xi_c}), were needed in the first-order expressions and hence also enter the inhomogeneities of the second order.

By introducing a Helmholtz decomposition also for the second-order flow,
\begin{align}
    \langle\vec{v}^{(2)}\rangle=\nabla\Phi^{(2)}+\nabla\times\vec{\Psi}^{(2)},\label{eq::Helmholtz_decomp_2nd}
\end{align}
the differential equations can be decoupled. 
First, we insert the decomposition 
Eq.~(\ref{eq::Helmholtz_decomp_2nd}) in the second order continuity equation, Eq.~(\ref{eq::second_order_contiuity}), to get
\begin{align}
    \nabla^2\Phi^{(2)}=-\frac{1}{\omega}\nabla\cdot\langle i k_f^2 \varphi^{(1)}\vec{v}^{(1)}\rangle.\label{eq::DGL_Phi}
\end{align}
Here, the right-hand side was simplified by using the first-order continuity of mass, Eq.~(\ref{eq::first_order_continuity}), to eliminate $\rho^{(1)}$, and by 
applying the Helmholtz decomposition to simplify the divergence of $\vec{v}^{(1)}$.

Second, we apply a rotation to Eq.~(\ref{eq::second_order_momentum_stress_inserted}) and insert Eq.~(\ref{eq::Helmholtz_decomp_2nd}), resulting in 
\begin{align}
    \nabla^4\vec{\Psi}^{(2)}=&-\frac{\rho^{(0)}}{\eta_f+\eta_p}\nabla\times\langle\vec{v}^{(1)}\nabla\cdot\vec{v}^{(1)}+\vec{v}^{(1)}\cdot\nabla\vec{v}^{(1)}\rangle\nonumber\\
    &-\frac{\lambda_M}{\eta_f+\eta_p}\nabla\times\langle\nabla\cdot\mathbf{T}\rangle,\label{eq::DGL_Psi}
\end{align}
where we used $\nabla\cdot\vec{\Psi}^{(2)}=0$ to simplify the result, which holds due to the axial symmetry. 

Note that the second term on the right-hand side is a new contribution due to the Oldroyd-B model, containing the convected derivative. 
The first term looks the same 
as for a simple viscous medium, only with viscosity $\eta=\eta_f+\eta_p$, which reflects the behavior of the Oldroyd-B model on long time scales. Care should be taken not to neglect the velocity divergence $\vec{v}^{(1)}\nabla\cdot\vec{v}^{(1)}$, even in the limit $\tfrac{\delta^2}{\lambda^2}\ll 1$, as the boundary conditions depend sensitively on the inhomogeneities, especially in cases where the inhomogeneities are strongly oscillating. 

Equations~(\ref{eq::DGL_Phi}) and (\ref{eq::DGL_Psi}) 
can be solved by expanding them in Legendre polynomials, similar to the first order,
\begin{align}
    \Phi^{(2)}(r,\theta)&=\sum_{l=0}^\infty \Phi^{(2)}_l(r)P_l(\mu),\label{eq::reihe_phi2}\\
    \vec{\Psi}^{(2)}(r,\theta)&=\hat{e}_\phi\sum_{l=1}^\infty \Psi^{(2)}_l(r)P_l^1(\mu),\label{eq::reihe_psi2}
\end{align}
which results in
\begin{align}
    \langle v^{(2)}_r\rangle(r,\theta)&=\sum_{l=0}^\infty v^{(2)}_{rl}(r)P_l(\mu),\label{eq::V2rReihe}\\
    \langle v^{(2)}_\theta\rangle(r,\theta)&=\sum_{l=1}^\infty v^{(2)}_{\theta l}(r)P_l^1(\mu),
\end{align}
for the Eulerian streaming velocity, with the radial functions given by
\begin{align}
    v^{(2)}_{rl}(r)&=\frac{\partial\Phi^{(2)}_l(r)}{\partial r}-\frac{l(l+1)}{r}\Psi^{(2)}_l(r),\label{eq::V2r}\\
    v^{(2)}_{\theta l}(r)&=\frac{\Phi^{(2)}_l(r)-\Psi^{(2)}_l(r)}{r}-\frac{\partial\Psi^{(2)}_l(r)}{\partial r}.\label{eq::V2theta}
\end{align}
The solutions to the radial function of the potentials are calculated separately in each medium using variation of constants and subsequently combined by imposing continuous tangential flow and stress as well as vanishing radial velocity at the boundary. Details of the calculation are provided in Appendix~\ref{app::phi}-\ref{app::BCs}.

\begin{table*}
\caption{Used Parameters}
\label{tab::parameters}
\begin{ruledtabular}
\begin{tabular}{lcccccccccc}
System &
\begin{tabular}{c}
$c_\mathrm{Particle}$ \\
$[\mathrm{m/s}]$
\end{tabular} &
\begin{tabular}{c}
$\rho_\mathrm{Particle}$ \\
$[\mathrm{kg/m^3}]$
\end{tabular} &
\begin{tabular}{c}
$\eta_\mathrm{f, Particle}$ \\
$[\mathrm{Pa\,s}]$
\end{tabular} &
\begin{tabular}{c}
$\eta_\mathrm{p, Particle}$ \\
$[\mathrm{Pa\,s}]$
\end{tabular} &
\begin{tabular}{c}
$\lambda_\mathrm{M, Particle}$ \\
$[\mathrm{s}]$
\end{tabular} &
\begin{tabular}{c}
$c_\mathrm{Medium}$ \\
$[\mathrm{m/s}]$
\end{tabular} &
\begin{tabular}{c}
$\rho_\mathrm{Medium}$ \\
$[\mathrm{kg/m^3}]$
\end{tabular} &
\begin{tabular}{c}
$\eta_\mathrm{f, Medium}$ \\
$[\mathrm{Pa\,s}]$
\end{tabular} &
\begin{tabular}{c}
$R$ \\
$[\mathrm{\mu m}]$
\end{tabular} &
\begin{tabular}{c}
$f$ \\
$[\mathrm{MHz}]$
\end{tabular} \\
\colrule
1) PEO solution in water & 1500 & 1000 & 0.001 & 0.009 & $10^{-6}$ & 1500 & 1000 & 0.001 & 2.5-25 & 0.01-1 \\
2) Model cell w/o contrast  & 1500 & 1000 & 0.001 & \cite{wozniak} & $10^{-6}$ & 1500 & 1000 & 0.001 & 15 & 0.01-1 \\
3) Model cell with contrast  & 1500 & 1100 & 0.001 & \cite{wozniak} & $10^{-6}$ & 1480 & 1000 & 0.001 & 15 & 0.01-1 \\
4) PEO solution in medium  & 1500 & 1000 & 0.001 & 0.009 & $10^{-6}$ & 1500 & 1000 & 0.001-0.1 & 15 & 0.01-1 \\
\end{tabular}
\end{ruledtabular}
\end{table*}

When investigating the streaming flow, it is more instructive to consider the Lagrangian velocity, which characterizes the particle transport. It  can be calculated by adding the Stokes drift to the Eulerian velocity \cite{10.1098/rspa.1998.0183},
\begin{align}
    \vec{V}_{L}=\langle\vec{v}^{(2)}\rangle+\left\langle\frac{i}{\omega}\vec{v}^{(1)}\cdot\nabla\vec{v}^{(1)}\right\rangle.\label{eq::app::stokesdrift}
\end{align}

\begin{figure*}[t!]
     \centering 	\includegraphics[width=0.95\linewidth, trim={0cm 0cm 0cm 0cm},clip,angle=0]{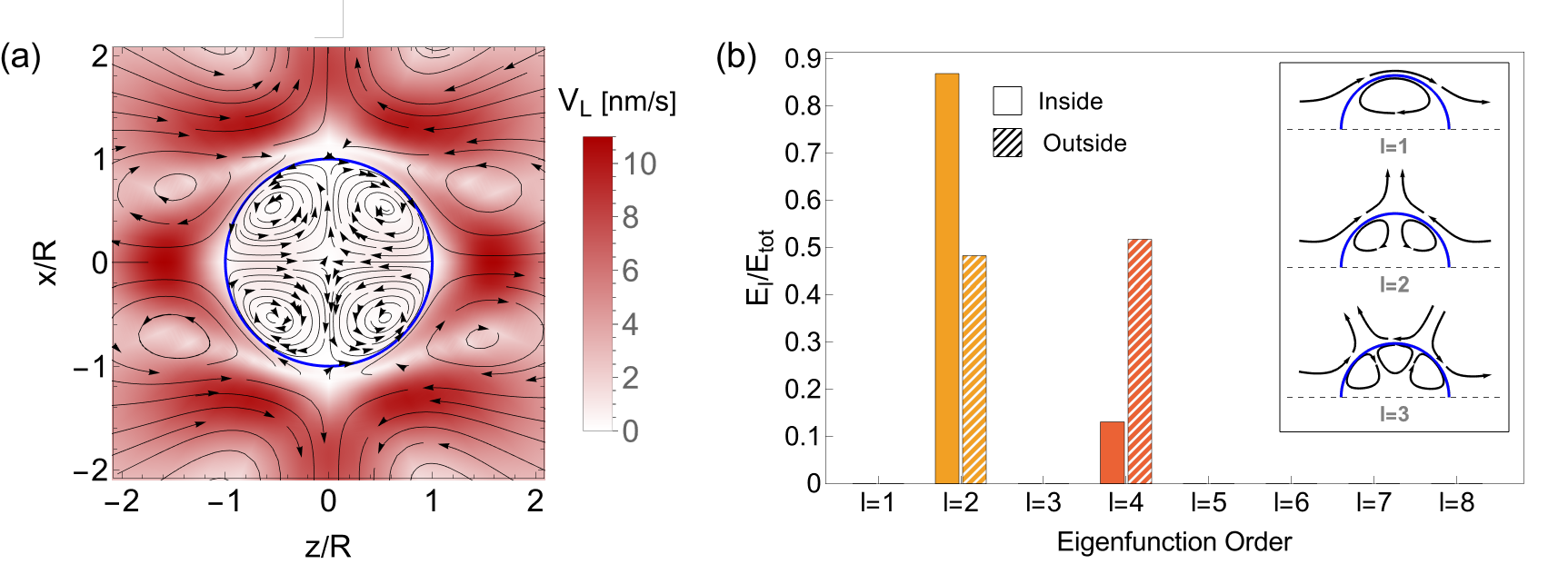}
 	    \caption{Energy decomposition into different eigenmodes. 
        (a) Example of a streaming flow field, with the absolute magnitude of the velocity shown in red.
        (b) Resulting kinetic energy density per eigenmode for the flow field shown in (a), normalized to the total kinetic energy density. The decomposition of the internal and external streaming flows 
        is shown as the filled and hashed bars, respectively. Values below
        one percent are not shown. The inset of (b) shows the characteristic streamline patterns for the first three eigenmodes.    \label{img::representation_energy}} 
\end{figure*}

With this, we have fully solved the streaming flows both inside and outside of the particle. Note that we have imposed no restriction on radius, viscous boundary layer thickness, or wavelength. In the following section we will show and discuss our results, which necessitate a swift convergence of the partial wave expansion to truncate the series, imposing a condition on the wavelength/wave number as $\Re(k_f)R\ll1$.

\section{Results}

\subsection{Streaming patterns}

While the inhomogeneities for each mode of Eqs.~(\ref{eq::reihe_phi2}) and (\ref{eq::reihe_psi2}) can, in principle, be derived analytically for arbitrary $l$, the derivation is unwieldy and offers only limited physical insight. Hence the inhomogeneities are calculated order-by-order using symbolic computation with \textit{Mathematica}, assuming the first order velocities are truncated at a specific maximal eigenmode $N_\mathrm{max}$. The resulting differential equations for the coefficients in the series expansion are then solved using variation of the constants, cf.~Eqs. (\ref{app::C1})-(\ref{app::C2}) and (\ref{app::C3})-(\ref{app::C6}). In theory these integrals can be solved analytically by expanding all terms into expressions of the form $e^{kr}/r^l$, which can be integrated using incomplete gamma functions. However, the expansion of the spherical Bessel functions into its oscillatory terms results in expressions that diverge for small arguments, which happens as long as $\Re(k_f)R\ll1$. The spherical Bessel functions themselves are well behaved due to precise cancellations. Therefore, the analytical solution is numerically less stable than direct numerical integration of the Bessel functions, and the analytical solution is only employed for the far field integrals in Appendix \ref{app::limitingbehaviour}. This semi-analytic approach was implemented in a \textit{Mathematica}-Script, available as Supplementary Material.

In the following, we chose the excitation to be
a plane standing wave with 
$p_\mathrm{ac}=1\,\mathrm{MPa}$. 
The particle is always placed in the 
middle between the node and the antinode of the standing wave. 
The expansion is truncated after the fourth eigenmode, which is deemed sufficiently converged for our parameters, where 
$\Re(k_f)R\lesssim0.1$.
The resulting streaming flows will then, due to the nonlinearity, have terms up to the eighth-order eigenmode. 
To simplify the discussion in the following, we use the term "order" 
only to refer to the orders of the acoustic expansion, and "mode" to refer to the orders of the eigenmode decomposition. 
In order to keep the parameter space tractable, 
we set the bulk viscosity for all media to zero, $\xi_p=\xi_f=0$. 
The other material parameters for all figures are provided in Table~\ref{tab::parameters}.

In order to visualize the parameter scans in the following, it is convenient and insightful to
introduce the average kinetic energy density (called energy density in the following):
\begin{align}
    E&=\frac{1}{V}\frac{1}{2}\rho^{(0)}\int_V\vec{V}_L(r,\theta)^2\mathrm{d}\vec{r}\nonumber\\
    &=\frac{1}{V}\pi\rho^{(0)}\sum_{l=1}^{N_\mathrm{max}}\left\{\int_{R_\mathrm{min}}^{R_\mathrm{max}}\left[\frac{2}{2l+1}V_{L,rl}(r)^2\right.\right.\nonumber\\
    &\hspace{3cm}\left.\left.+\frac{2l(l+1)}{2l+1}V_{L,\theta l}(r)^2\right]r^2\mathrm{d}r\right\},
\end{align}
where we introduced the spherical components of the Lagrangian velocity,
cf.~Eq.~(\ref{eq::app::stokesdrift}).
Due to the quadratic dependence on the streaming velocity, the total energy density decomposes into a sum of the energy densities contained in each eigenmode. Here, the reference volume depends on whether the energy density inside or outside is considered. 
The inner energy density is averaged over the volume of the sphere  ($R_\mathrm{min}=0$ and $R_\mathrm{max}=R$), while the outer energy density is averaged over the computational domain (chosen from $R_\mathrm{min}=R$ and $R_\mathrm{max}=3R$). As we are predominantly interested in the streaming on the scale of the fluid droplet, the computational domain outside is chosen with respect to the droplet radius, even though the wave length of the compressional wave is much larger.

An example of this energy decomposition 
is shown in Fig.~\ref{img::representation_energy}. Panel (a) shows an example streaming flow, where the magnitude of the velocity is shown in red, overlayed with the streamlines of the flow. Its energy decomposition is represented in panel (b), which shows the average kinetic energy density per mode normalized to the total energy for both internal and external flows in full and hashed bars, respectively. 
First, note that each eigenmode has a characteristic streamline pattern as shown in the inset of Fig.~\figref{img::representation_energy}{(b)} (due to the axial symmetry of the problem, only the upper half-space is shown). 
The overall structure of the flow is, therefore, well represented by the relative strength of these modes, hence the energy density, cf. Fig.~\figref{img::representation_energy}{(a)} and ~\figref{img::representation_energy}{(b)}. 
In this example,
the internal streaming in Fig.~\figref{img::representation_energy}{(a)} 
is in shape close to the characteristic streamlines of the second mode, which is reflected in $E_2$, shown by the solid yellow bar, being the dominant term in the energy density. 
The external flow exhibits four lobes, which however are significantly deformed and compressed, which is captured by the second and fourth mode having comparable strength as indicated by the hashed bars. 
Note that the modes $l>4$ have an incomplete leading order due to the truncation of the first-order solution.
However, as the expansion converges and their contributions are negligible, 
as seen in Fig.\figref{img::representation_energy}{b}, this does not pose a problem. 
Therefore, we show only the first four modes in the following discussions.
%  Note that due to the quadratic nonlinearity, the resulting second order is essentially formed by Clebsch-Gordan coefficients. As the first order converges, the resulting leading order contribution to each mode is given by the lowest possible combination of first-order contributions present in that mode. Due to the truncation of the first-order solution, the modes $l>3$ shown here have an incomplete leading order.

While the chosen normalization of the energy density per mode is intuitive, 
it is not suitable to compare different simulations, where the total energy density changes. Therefore, in the following, we will rather normalize the kinetic energy density by the acoustic energy density contained in the standing wave, assuming for simplicity an inviscid medium, i.e.~using $E_\mathrm{ac}=\frac{p_\mathrm{ac}^2}{4\rho^{(0)}c^2}$.

\begin{figure*}[t]
     \centering
 \includegraphics[width=0.95\linewidth, trim={0cm 0cm 0cm 0cm},clip,angle=0]{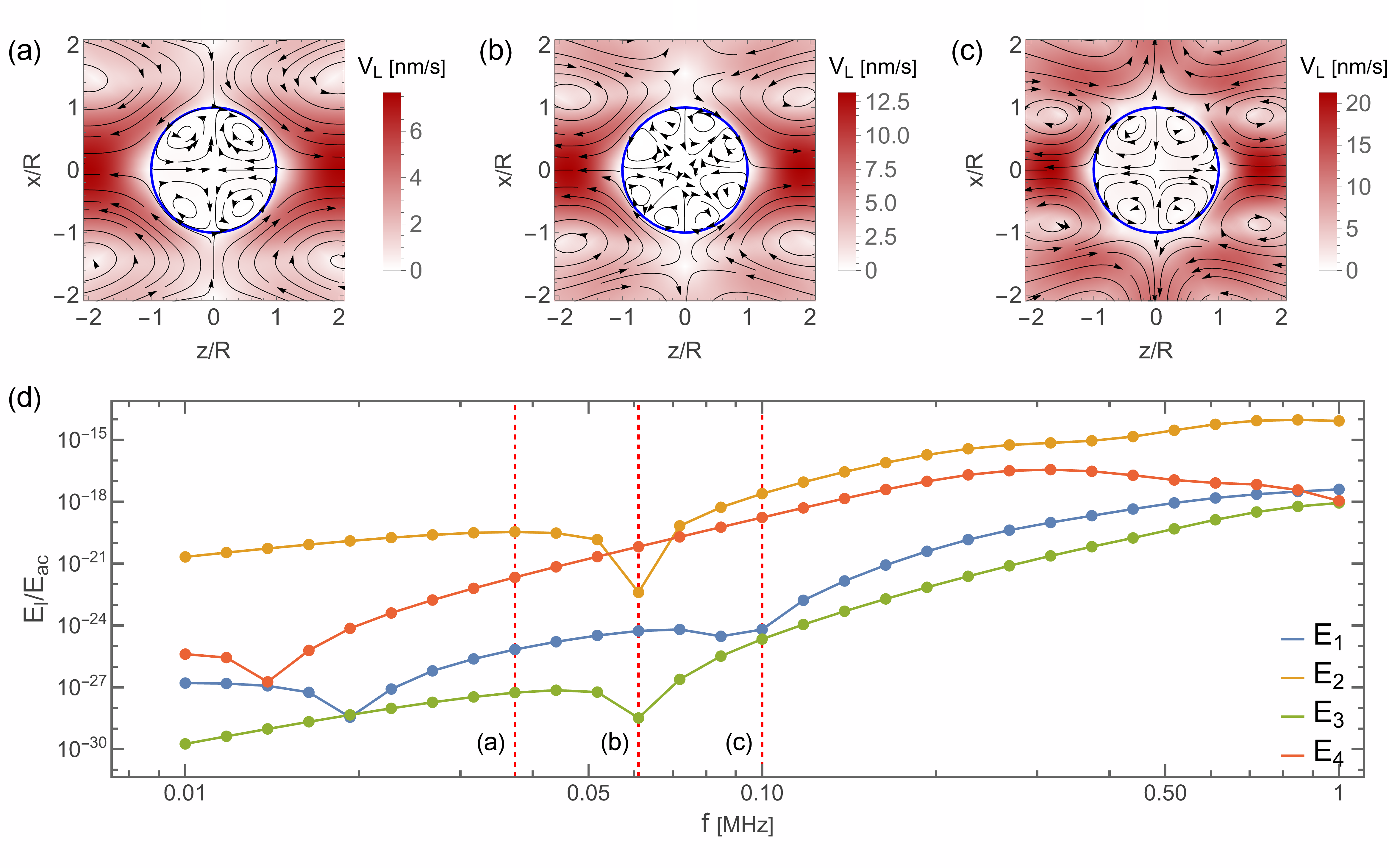}
 	    \caption{Flow reversal in a spherical PEO-droplet 
        in water. (a)-(c) Representative streaming flows for three different frequencies. (d) Energy density of the first four modes inside the sphere, normalized to the energy density of the acoustic wave, as a function of the frequency. The red vertical lines denote the respective frequencies of the streaming flows shown in (a)-(c).
        Used parameters $\eta_f=0.001\,\mathrm{Pa~s}$, $\eta_p=0.009\,\mathrm{Pa~s}$, $\lambda_M=10^{-6}\,\mathrm{s}$, $R=4.09\,\mathrm{\mu m}$.
        \label{img::flowreversal_line}} 
\end{figure*}

\subsection{Droplet of PEO solution}

As a first benchmark problem,
we investigated a spherical droplet
of aqueous polyethylen oxide (PEO, also known
as polyethylene glycol, PEG) solution suspended 
in water as the surrounding medium, which is a standard
choice to tune the viscosity of an aqueous solution
in biomedical applications. For reasons of comparison,  
we used the numerical parameters from Doinikov \textit{et al.} \cite{VS_Doinikov} for the PEO solution: $\eta_f=0.001\,\mathrm{Pa~s}$, 
$\eta_p=0.009\,\mathrm{Pa~s}$, and 
$\lambda_M=10^{-6}\,\mathrm{s}$ with the
remaining parameters given as system 1 in Table \ref{tab::parameters}. 

Figure~\ref{img::flowreversal_line} shows the resulting internal energy density per mode normalized to the acoustic energy density of the incident wave, 
for a droplet radius of $R=4.09\,\mathrm{\mu m}$ and for different frequencies $f=\frac{\omega}{2\pi}$,
in a log-log scale. Additionally, three representative flow fields at specific frequencies are shown. The maximum streaming velocities for the representative flow fields are on the order of several nanometers per second, however, for higher frequencies values over $0.1\,\mathrm{\mu m}$ are reached.
Note that in contrast to what is  
typically assumed for 
calculations of the acoustic radiation force \cite{C2LC21068A}, there is no density or compressibility contrast between the two media here. 
Instead, to focus on the rheological model, the scattering is purely mediated by the difference in viscosity and hence is lower than typical.

Generally, the energy density is dominated by the second mode, 
which is proportional to the quadrupolar field,
as shown by the yellow curve in 
Fig.~\figref{img::flowreversal_line}{(d)}. However, around $60\,\mathrm{kHz}$ the contribution of the second mode collapses and the response is dominated by the fourth mode instead, shown in red. 
This is clearly reflected 
in the streamlines of the three example flows (with their respective frequencies denoted by the red dashed lines): 
The internal flow has a purely quadrupolar character for Figs.~\figref{img::flowreversal_line}{(a)}~and~\figref{img::flowreversal_line}{(c)}, and a purely hexadecapolar ($l=4$) character in Fig.~\figref{img::flowreversal_line}{(b)},
where the second mode collapsed. Because of the good correspondence between the flow profiles and the energy density, we will not show the explicit flows in the following and focus on the energy representation. Note that the energy density plots shown here and in the following only encode the internal field and not the external one.

It should be noted that in Fig.~\figref{img::flowreversal_line}{(b)}~and~~\figref{img::flowreversal_line}{(c)}, the internal and external flow field do not seem to match up regarding their flow direction, which would violate the boundary condition of continuous tangential Lagrangian flow. Instead, while the tangential stress is continuous on the boundary, the Eulerian velocity gradient can have a significant discontinuity due to the contribution of the first order stress on the oscillating boundary (cf.~Eq.~\ref{eq::app::2ndtangential_stress_BC}) resulting in very steep gradients, while still maintaining the prescribed boundary conditions.

Strikingly, the flow direction of the streamlines reverses upon crossing the 
sharp drop of the second mode of the energy density. 
This can be seen by comparing the direction of the flow inside, which is flowing toward $z=0$ along the $z$-direction in Fig.~\figref{img::flowreversal_line}{(a)} 
and away from $z=0$ in Fig.~\figref{img::flowreversal_line}{(c)}.

Note that all other modes show similar sharp drops. 
A thorough investigation of these changes in energy distribution and flow reversals is performed next, using a more condensed representation to be able to study larger parameter regions.

\begin{figure*}[t!]
     \centering
 	\includegraphics[width=1\linewidth, trim={0cm 0.5cm 0cm 0cm},clip,angle=0]{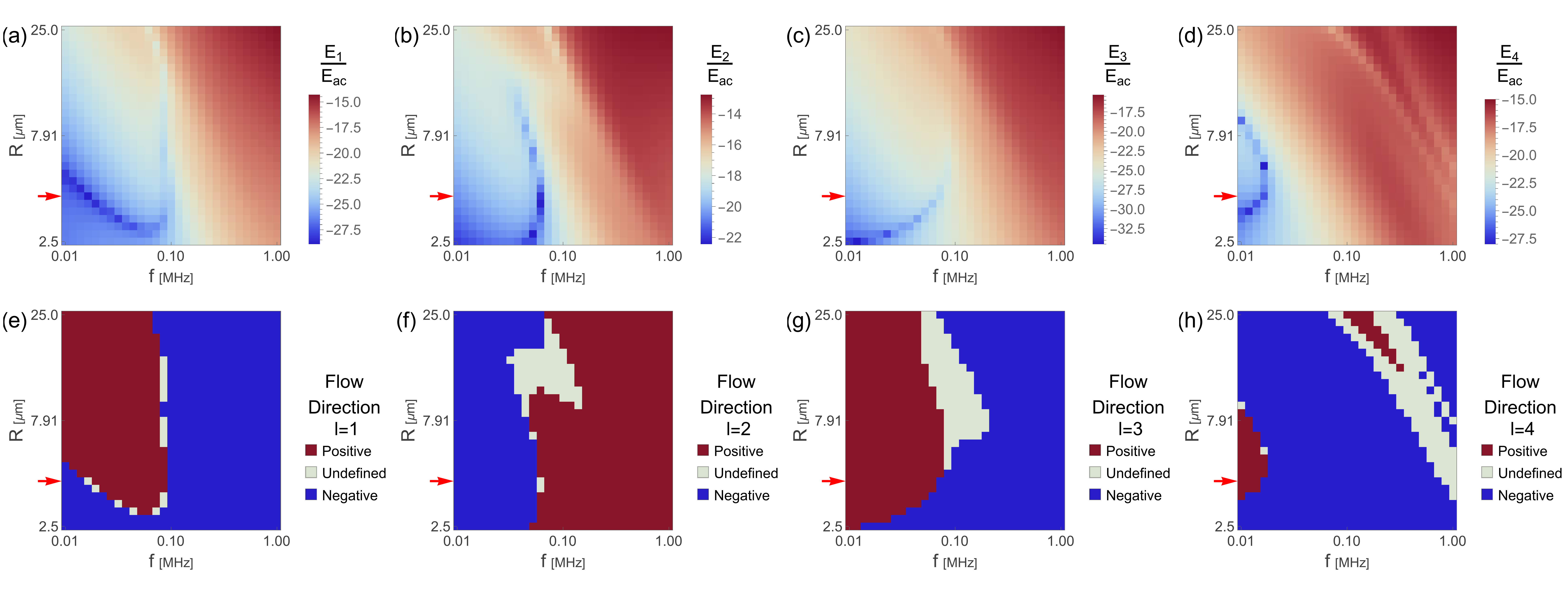}
 	    \caption{Flow reversals inside a spherical PEO-droplet in water, as a function of droplet radius and frequency. The internal energy density per mode normalized to the acoustic energy density is shown in the top row (a)-(d), the radial flow direction for each eigenmode is shown in the second row (e)-(h). 
        Droplet parameters: $\eta_f=0.001\,\mathrm{Pa~s}$, $\eta_p=0.009\,\mathrm{Pa~s}$, $\lambda_M=10^{-6}\,\mathrm{s}$.
        \label{img::PEO}} 
\end{figure*}

Figure~\ref{img::PEO} shows the same PEO in water system in a 2D parameter scan 
over both the droplet radius $R$ and 
the frequency $f$. 
The first row, Fig.~\figref{img::PEO}{(a)-(d)} shows the energy density for the first four modes. 
Note that the color scales are different 
for all four plots and logarithmic. The small red arrow at each plot indicates the row which corresponds to Fig.~\ref{img::flowreversal_line}. 
The bottom row Fig.~\figref{img::PEO}{(e)-(h)} shows the flow direction of the internal streaming, as calculated from determining the overall sign of 
the Lagrangian velocity componets $V_{L,rl}(r)$ 
between $0$ and $R$, 
which is equivalent to the streaming flow along the positive $z$-axis. 
The cases marked as "undefined", shown in grey, relate to a sign flip of the velocity along the radial direction, which suggests the existence of radial steamrolls in the flow, as checked by visual inspection of the
flow fields.

\begin{figure*}[t!]
     \centering
 \includegraphics[width=1\linewidth, trim={0cm 0.5cm 0cm 0cm},clip,angle=0]{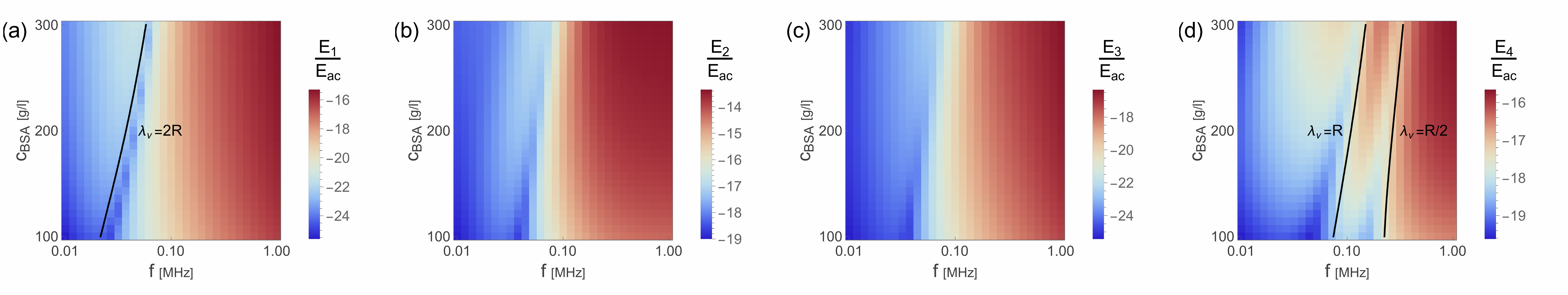}
 	    \caption{The internal kinetic energy density per eigenmode of the streaming flows in a model cell without density and velocity contrast suspended 
        in water. The black lines in the $l=1$ and $l=4$ mode in (a) and (d) denote where the wavelength of the shear wave equals the droplet diameter, radius, or half the radius, respectively. 
        Droplet parameters: $R=15\,\mathrm{\mu m}$, 
        $\lambda_M=10^{-6}\,\mathrm{s}$, $\eta_f=0.001\,\mathrm{Pa~s}$, 
        $\eta_p$ modeled using the data of \cite{wozniak}. 
      \label{img::Cell}} 
\end{figure*}

Comparing  Fig.~\figref{img::PEO}{(a)-(d)}
and  Fig.~\figref{img::PEO}{(e)-(h)},
for each of the four modes, the lines at which the energy density of the respective mode drops coincide with a 
reversal of the flow direction. 
These flow reversals are specific 
for each mode and can occur either directly or through a transition area during which radial steamrolls flip the direction of the flow. 
Because of this clear correspondence, we will not explicitly show the flow direction plots in the following, 
and identify the lines of low energy density with lines of flow reversal. 

It should be noted 
that the functional shape of 
the lines of flow reversal in the $R$-$f$-plane are not simple 
and furthermore differ for each mode. This shows that the behavior 
is not just governed by the scattering parameter $\Re(k_f)R$, but by a
complex nonlinear interaction of multiple contributions. For example, the frequency not only influences  the wavelength of the compression and shear waves differently, cf.~Eqs.~(\ref{eq::kf}) and (\ref{eq::knu}), but also influences the first-order material parameters via Eqs.~(\ref{eq::eta_c}) and (\ref{eq::app::tau1}). These first-order expressions then interact nonlinearly in the inhomogeneity in Eq.~(\ref{eq::DGL_Psi}).

\subsection{Protein solution as model for suspended cell}

While having a well-established rheology,
to describe intracellular streaming
a polymer solution such as
PEG is not sufficient.
Instead, we now consider a model cell 
composed of an aqueous solution 
of (globular) proteins.
As a well-characterized protein,
we consider a droplet of  
an aqueous solution of
bovine serum albumin (BSA) 
with a droplet radius of $15\,\mathrm{\mu m}$, corresponding to a typical cellular size, suspended in water. 
Note that we use the BSA as a stand-in 
for the complex collection of proteins present in a suspended cell (such as globular actin, etc).

Following the study of Woznaik \textit{et al.} \cite{wozniak}, the viscosity of the BSA solution is well modeled using a quadratic generalization of the Einstein relation for 
hard spheres $\eta_r=1+2.5\Xi+b\Xi^2$, where $\eta_r$ is the viscosity relative to the solvent, typically water,
and $\Xi$ is the volume fraction of the solution. We take their result 
of $b=93$ and convert the concentration to the volume fraction by 
$c[\mathrm{g/l}]=1430\,\Xi$ 
to investigate a concentration between $100\,\mathrm{g/l}$ and $300\,\mathrm{g/l}$, which is in line with the concentration of macromolecules reported in biological cells \cite{ZIMMERMAN1991599}. 
Note that the viscosity of Wozniak \textit{et al.} is equivalent to the long-time limit of the Oldroyd-B model, therefore $\eta_p=\eta_w(2.5\Xi+b\Xi^2)$ where $\eta_w$ is the viscosity of water.
For the viscoelastic time scale, we use 
$\lambda_M=10^{-6}\,\mathrm{s}$. While the timescale is expected to also change with the concentration, this effect is considered less important and has not been thoroughly studied so far.

To study the impact of the Oldroyd-B rheology, we again consider
the case of no density or compressibility contrast first. The importance of these
additional effects is then investigated in a second step. All used material parameters can be found as system 2
in Table \ref{tab::parameters}.

\begin{figure*}[t!]
     \centering
 \includegraphics[width=1\linewidth, trim={0cm 0.5cm 0cm 0cm},clip,angle=0]{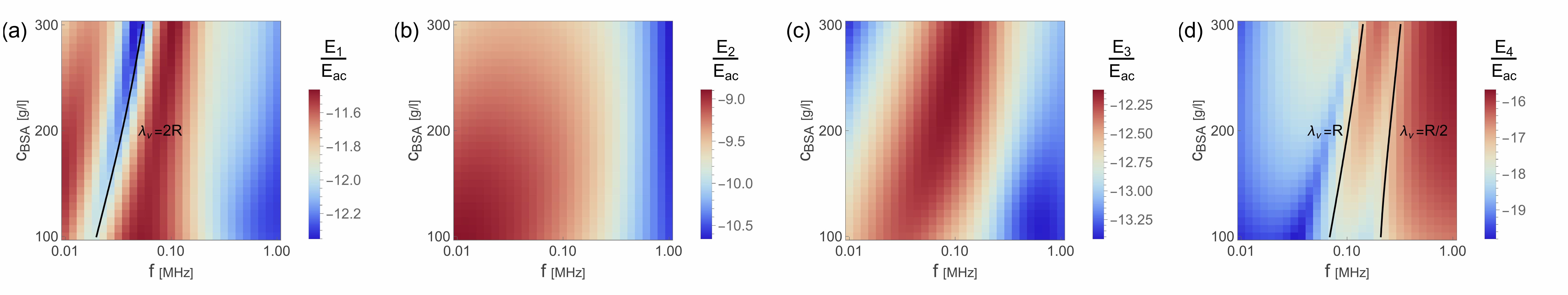}
 	    \caption{The internal kinetic energy density per eigenmode of the streaming flows in a model cell with density and velocity contrast
        suspended in water, cf. system 3 in Table \ref{tab::parameters}. The black lines in the $l=1$ and $l=4$ mode in (a) and (d) denote where the wavelength of the shear wave equals the droplet diameter, radius, or half the radius, respectively.
        Droplet parameters: $R=15\,\mathrm{\mu m}$, $\lambda_M=10^{-6}\,\mathrm{s}$, 
        $\eta_f=0.001\,\mathrm{Pa~s}$, 
        $\eta_p$ modeled using the data of \cite{wozniak}.
    \label{img::Celldensity}} 
\end{figure*}

Figure~\ref{img::Cell} shows
the average internal energy densities per mode as a function of both the  BSA concentration and the driving frequency.
Similar to the streaming in the case of a  PEO droplet, 
the response is generally dominated by the second mode, with the third mode having the smallest contribution. 
There is a very clear line of flow reversal in the first three modes below 100 kHz. While the transition shifts towards higher frequencies with increasing mode, a second transition appears in the fourth mode in Fig.~\figref{img::Cell}{(d)} at some hundreds of kHz.
The transitions of the flow direction occur at higher frequencies the higher the protein density, and thus the higher the polymer viscosity. Interestingly, for the first and fourth mode, the transitions occur roughly at the
frequency where the wavelength of 
the shear wave 
$\lambda_\nu=\frac{2\pi}{\Re(k_\nu)}$ fits once, twice or four times in the diameter of the droplet, respectively. This conditions are shown in Fig.~\figref{img::Cell}{(a)}~and~\figref{img::Cell}{(d)} as black lines. 
Note that the shape of the lines of flow reversal for the second and third mode in Fig.~\figref{img::Cell}{(b)}~and~\figref{img::Cell}{(c)} do not agree with the wavelength of the shear wave, so the full behavior of the second-order streaming response is still complex and determined by the full inhomogeneity.

To investigate their impact on the streaming, we also 
performed a parameter study 
where both density and sound velocity 
contrast are considered. For this,
following \cite{C2LC21261G},
we assume for the protein solution
$\rho_\mathrm{Particle}=1100\,\mathrm{kg/m^3}$
and $c_\mathrm{Particle}=1500\,\mathrm{m/s}$, and adjust the speed of sound in water to $c_\mathrm{Medium}=1480\,\mathrm{m/s}$ to reflect the added precision, cf. system 3 in Table \ref{tab::parameters}.
For comparability, we performed 
the same parameter scan as in Fig.~\ref{img::Cell}. The resulting
energy densities per mode are shown in Fig.~\ref{img::Celldensity}.

The fourth mode in Fig.~\figref{img::Celldensity}{(d)} stays almost the same between both models, with only minor shifts in the position of the flow reversals. For the other three modes, the changes are drastic. The maximal values of the energy density increase by roughly four orders of magnitude and there are significant changes in the qualitative behavior. While the first mode shown in Fig.~\figref{img::Celldensity}{(a)} retains its flow reversal, it is shifted to lower frequencies, and the energy density falls off with rising frequencies after the reversal. Both the second and third mode no longer exhibit any flow reversals. 
Strikingly, the second mode  shown in Fig.~\figref{img::Celldensity}{(b)} no longer increases, but now decreases with rising frequency.

The trend of the first and second mode suggest that for frequencies above $1\,\mathrm{MHz}$ the first mode might replace the second mode as the dominant contribution, as also seen in \cite{Thierry_SiB}, for purely viscous materials and different density contrasts. 
The large increase in the kinetic energy density stems from the increased scattering due to the stronger contrast and underlines that while 
for a cell-like protein solution
the density contrast is just $10\%$ and 
the sound velocity contrast 
below $2\%$ , 
their contribution to the streaming flows can be dominant over the viscosity contrast and should not be neglected.

\begin{figure*}[t!]
     \centering
 \includegraphics[width=1\linewidth, trim={0cm 0.5cm 0cm 0cm},clip,angle=0]{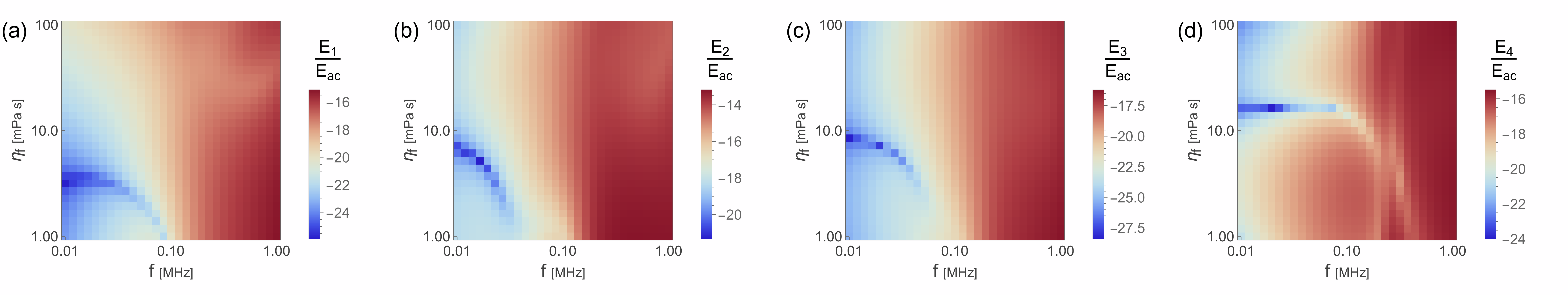}
 	    \caption{The internal kinetic energy density per eigenmode of the streaming flows in a sphere of PEO solution immersed in external media 
        of variable viscosity $\eta_f$, as a function of medium viscosity
        and frequency.
        Droplet parameters: $\eta_f=0.001\,\mathrm{Pa~s}$, 
        $\eta_p=0.009\,\mathrm{Pa~s}$, 
        $\lambda_M=10^{-6}\,\mathrm{s}$.
        \label{img::Sensing}} 
\end{figure*}

\subsection{Possible application for sensing}

An additional possible application of our theory, beyond quantitatively assessing streaming in suspended cells,
originates in the fact that the 
predicted flow reversals 
occur separately for each mode. 
This suggests the possibility to
classify a parameter space by the specific combination of the flow directions for every mode. 
This idea could be applied in a sensing method, where either the internal or external flow is measured using particle image velocimetry, 
and a spatial multipole decomposition is carried out to separate the different modes. 
Depending on the setup, either 
the viscosity of the droplet 
or of the surrounding medium could be determined, provided the other material is well known. This could be used either to determine the viscosity dispersion between multiple droplets or as a micro-rheometer for the external medium, if it is difficult to obtain macroscopic amounts of the sample. 

To investigate this proposition, we simulated the streaming inside 
a PEO droplet with a radius of 
$R=15\,\mathrm{\mu m}$ when the viscosity of the external medium is changed. For this, we still assume a Newtonian fluid for the external medium and modify its fluid viscosity from $1-100\,\mathrm{mPa~s}$, 
while again assuming no density or compressibility contrast. 
The detailed material parameters are given as system 4 in Table \ref{tab::parameters}.

The resulting average energy densities per mode are shown in Fig.~\ref{img::Sensing} as a function of both the frequency and the viscosity of the medium. Note that we show the internal energy density here. 
However, the same arguments could be made for the external energy density if that is more suitable for the application in mind.
In the example shown in Fig.~\ref{img::Sensing}, considering small frequencies, the four modes switch flow directions for increasing viscosity, and hence from their specific directions alone one can estimate the latter within the shown viscosity range. Additionally, while the flow reversals for the four modes show some similarities, they do not coincide, which allows classifying the viscosity more precisely using the combination of the direction of flow of each mode.

\section{Conclusion}

In this paper, we have developed a semi-analytic method to describe the streaming inside and outside of a spherical particle or droplet, when both media are described by an Oldroyd-B fluid. 
While the full analytical treatment considers all multipolar orders and imposes no restrictions on the thickness of the viscous boundary layer or the wavelength compared to the particle size, the numerical implementation requires truncation after a set number of multipolar orders, which only converges if the scattering parameter is small ($\Re(k_f)R\ll1$).

We applied our method first to PEO droplets in viscous media, where we showed that the overall shape of the streamlines can change dramatically when varying the material parameters. 
Furthermore, we investigated the streaming in a model cell, containing a protein solution, where we showed that both the rheological as well as the density and compressibility contrast strongly influence the streaming. Specifically, albeit density and compressibility contrast are known to be quite small for cells (or cell
surrogates like hydrogels), they nevertheless allow for significant increase of the streaming magnitude
and can even strongly change the flow patterns qualitatively. These transitions demonstrate that
acoustofluidics is a highly non-linear problem; this can be a nuisance, because small
microscopic heterogeneities can have large macroscopic effects, but also an advantage, 
because small molecular differences can be amplified to the level of cells.

Finally, we proposed that classifying the flow direction of each multipolar order 
could be used to measure the rheological properties of either the droplet or the surrounding material.
This suggests that cells might use simple readouts of direction, e.g. by
sensory cilia \cite{cammann2025form}, to infer material properties of their environment, when
subjected to ultrasound stimulation.

In order to model cells more accurately, a multiscale rheological model should be considered. Specifically, the first and second order solutions probe the cell on different wavelengths. While it is valid to treat the cytoplasm as a dilute polymer solution for the acoustic waves, 
the streaming flows in the entire cell would be more accurately depicted using porous flow and/or poro-viscoelasticity.
Furthermore, the effect of the non-negligible cell surface tension on the boundary conditions should be investigated, as well as deformability in case of sufficiently strong flows. Another avenue
for future extensions is the inclusion of a mechanical model for the nucleus, which
often is modelled as an elastic inclusion \cite{wohlrab2024mechanical,chojowski2024role}.

When applying very general physical stimuli such as ultrasound to cells, one has to be careful about
other processes that might be triggered as well, including e.g.\ rupture of
cell membranes, cytoskeletal structures or adhesions. Another unwanted side effect
might be heating, which could lead to e.g.\ protein denaturation. 
Although such side effects might occur, they might be offset by
the advantage that very well defined flow patterns might be generated
within cells, offering a cheap and effective mean to control cell behaviour
by physical means.

\begin{acknowledgments}
The authors thank Athanasios Athanassiadis and Peer Fischer for helpful discussions. This work was supported by the Deutsche Forschungsgemeinschaft (DFG, German Research Foundation) via the Cluster of Excellence
3D Matter Made to Order (3DMM2O) (EXC-2082/1-390761711 and EXC-2082/2-390761711), as
well as by the Carl-Zeiss-Foundation.
\end{acknowledgments}

\appendix

\begin{widetext}

\section{Scattering Solution of the First Order}\label{App::scattering}
The scattering coefficients are calculated by applying boundary conditions to the first-order velocity and stress on the surface of the particle, given by
\begin{align}
    \vec{v}_\mathrm{o}^{(1)}=\vec{v}_\mathrm{ac}^{(1)}+\vec{v}_\mathrm{sc}^{(1)}&=\vec{v}_\mathrm{i}^{(1)},\label{eq::first_order_BC_velo}\\
    s_\mathrm{o}^{(1)}=s_\mathrm{ac}^{(1)}+s_\mathrm{sc}^{(1)}&=s_\mathrm{i}^{(1)},\label{eq::first_order_BC_normal_stress}\\
    t_\mathrm{o}^{(1)}=t_\mathrm{ac}^{(1)}+t_\mathrm{sc}^{(1)}&=t_\mathrm{i}^{(1)},\label{eq::first_order_BC_tangential_stress}\quad\mathrm{at}\quad r=R
\end{align}
where the indices $\mathrm{i}$ and $\mathrm{o}$ denote the solution inside and outside the sphere, respectively, and $t^{(1)}$ and $s^{(1)}$ are the components of the stress tensor tangential or normal to the surface. Their expressions 
\begin{align}
   t^{(1)}&= \eta_c\left(\frac{1}{r}\frac{\partial v_{r}^{(1)}}{\partial\theta}-\frac{v_{\theta}^{(1)}}{r}+\frac{\partial v_{\theta}^{(1)}}{\partial r}\right)\label{eq::first_order_tangential_stress},\\
    s^{(1)}&=\left(-\frac{ic^2\rho^{(0)}k_f^2}{\omega}\varphi^{(1)}+\left(\xi_c-\frac{2}{3}\eta_c\right)\left(\nabla\cdot\vec{v}^{(1)}\right)+2\eta_c\frac{\partial v_{r}^{(1)}}{\partial r}\right),\label{eq::first_order_normal_stress}
\end{align}
can be calculated by splitting the first-order stress into its components. Note that the normal component here refers only to the terms of $\hat{r}\otimes\hat{r}$, 
with $\otimes$ the dyadic product,
as we are interested in the component normal to a spherical particle. Additionally, the first order pressure in the normal stress was re-expressed as $\varphi^{(1)}$ using the adiabatic relation $p^{(1)}=c^2\rho^{(1)}$ before using the first order continuity  equation
\begin{align}
    \frac{\partial \rho^{(1)}}{\partial t}+\rho^{(0)}\nabla\cdot\vec{v}^{(1)}&=0\label{eq::first_order_continuity}.
\end{align}
This results in the following system of equations
\begin{align}
    m_{11}a_n+m_{12}b_n+m_{13}\hat{a}_n+m_{14}\hat{b}_n&=n_1\\
    m_{21}a_n+m_{22}b_n+m_{23}\hat{a}_n+m_{24}\hat{b}_n&=n_2\\
    m_{31}a_n+m_{32}b_n+m_{33}\hat{a}_n+m_{34}\hat{b}_n&=n_3\\
    m_{41}a_n+m_{42}b_n+m_{43}\hat{a}_n+m_{44}\hat{b}_n&=n_4
\end{align}
for each mode $n$, where the coefficients are given by
\begin{align}
\begin{aligned}
    m_{11}&=x_0h_n^{(1)'}(x_0),&
    m_{12}&=-n(n+1)h_n^{(1)}(x_{\nu 0}),&
    m_{13}&=-\hat{x}_0j_n'(\hat{x}_0),&
    m_{14}&=n(n+1)j_n(\hat{x}_{\nu 0}),\\
    m_{21}&=h_n^{(1)}(x_0),&
    m_{22}&=-h_n^{(1)}(x_{\nu 0})-x_{\nu 0}h_n^{(1)'}(x_{\nu 0}),&
    m_{23}&=-j_n(\hat{x}_0),&
    m_{24}&=j_n(\hat{x}_{\nu 0})+\hat{x}_{\nu 0}j_n'(\hat{x}_{\nu 0}),
\end{aligned}
\nonumber\\
\begin{aligned}
    m_{31}&=x_0^2[2\eta_c h_n^{(1)''}(x_0)-(i\rho^{(0)}c^2/\omega+\xi_c-2\eta_c/3)h_n^{(1)}(x_0)],&
    m_{32}&=2n(n+1)\eta_c[h_n^{(1)}(x_{\nu 0})-x_{\nu 0}h_n^{(1)'}(x_{\nu 0})],\\
    m_{33}&=\hat{x}_0^2[(i\hat{\rho}_0\hat{c}^2/\omega+\hat{\xi}_c-2\hat{\eta}_c/3)j_n(\hat{x}_0)-2\hat{\eta}_cj_n''(\hat{x}_0)],&
    m_{34}&=2n(n+1)\hat{\eta}_c[\hat{x}_{\nu 0}j_n'(\hat{x}_{\nu 0})-j_n(\hat{x}_{\nu 0})],\\
    m_{41}&=2\eta_c[x_0h_n^{(1)'}(x_0)-h_n^{(1)}(x_0)],&
    m_{42}&=-\eta_c[(n^2+n-2)h_n^{(1)}(x_{\nu 0})+x_{\nu 0}^2h_n^{(1)''}(x_{\nu 0})],\\
    m_{43}&=2\hat{\eta}_c[j_n(\hat{x}_0)-\hat{x}_0j_n'(\hat{x}_0)],&
    m_{44}&=\hat{\eta}_c[(n^2+n-2)j_n(\hat{x}_{\nu 0})+\hat{x}_{\nu 0}^2j_n''(\hat{x}_{\nu 0})],\\
    n_1&=-x_0j_n'(x_0)A_n,&
    n_2&=-j_n(x_0)A_n,\\
    n_3&=x_0^2[(i\rho^{(0)}c^2/\omega+\xi_c-2\eta_c/3)j_n(x_0)-2\eta_c j_n''(x_0)]A_n,&
    n_4&=2\eta_c[j_n(x_0)-x_0j_n'(x_0)]A_n,\\    
\end{aligned}
\end{align}
where a prime denotes a derivative with respect to the argument of the function, and
\begin{align}
    x_0&=kR,&
    x_{\nu 0}&=k_\nu R,&
    \hat{x}_0&=\hat{k}R,&
    \hat{x}_{\nu 0}&=\hat{k}_\nu R.
\end{align}

Note that for $n=0$, $b_n$ and $\hat{b}_n$ are zero by their definition Eqs.~(\ref{eq::expansion_b_n}) and (\ref{eq::expansion_b_n_hat}). Similarly, the equations for the tangential velocity and stress decompose in associated Legendre polynomials $P_n^1(\mu)$, and are therefore not defined in this case. As a result, for $n=0$ the problem reduces to two equations for two unknowns.

\section{Solution for $\Phi$\label{app::phi}}
The next sections follow the derivation given in \cite{VS_Doinikov} and are shown here in a compact form for convenience. We start with Eq.~(\ref{eq::DGL_Phi}), which can be represented by
\begin{align}
    \nabla^2\Phi^{(2)}=\frac{1}{2\omega}\Re\left[ik_f^{2*}\left(\frac{1}{r^2}\frac{\partial}{\partial r}(r^2\varphi^{(1)*}v_r^{(1)})+\frac{1}{r \sin\theta}\frac{\partial}{\partial \theta}(\varphi^{(1)*}v_\theta^{(1)}\sin\theta)\right)\right],
\end{align}
where we used $\langle fg\rangle=\frac{1}{2}\Re\left(f^*g\right)$ with $\Re$ denoting the real part, which holds for two periodic functions $f$ and $g$ with the same frequency. 

By expanding the inhomogeneity on the right-hand side, which we will call $f(r,\mu)$, in Legendre polynomials
\begin{align}
f(r,\mu)&=\sum_{l=0}^\infty\alpha_l(r)P_l(\mu),\label{eq::app::f(r)}\\
\alpha_l(r)&=\frac{2l+1}{2}\int_{-1}^1f(r,\mu)P_l(\mu)\mathrm{d}\mu,\label{eq::app::alpha}
\end{align}
the partial differential equation can be reduced to only one variable. The expansion can be performed explicitly using Clebsch-Gordan Coefficients \cite{VS_Doinikov}; in our 
semi-analytical treatment, 
the expansion is performed using \textit{Mathematica}. When inserting the expansions Eqs.~(\ref{eq::reihe_phi2}) and (\ref{eq::app::f(r)}) in Eq. (\ref{eq::DGL_Phi}) one arrives at
\begin{align}
    \frac{\partial^2\Phi^{(2)}_l(r)}{\partial r^2}+\frac{2}{r}\frac{\partial \Phi^{(2)}_l(r)}{\partial r}-\frac{l(l+1)}{r^2}\Phi^{(2)}_l(r)=\alpha_l(r),
\end{align}
which can be solved by variation of constants, yielding the solution
\begin{align}
    \Phi^{(2)}_l(r)=\frac{C_{1l}(r)}{r^{l+1}}+C_{2l}(r)r^l.\label{eq::app::Phi_sol}
\end{align}
Here the coefficients are determined by
\begin{align}
    C_{1l}(r)=C_{1l0}-\frac{1}{2l+1}\int_R^r s^{l+2}\alpha_l(s)\mathrm{d}s,\label{app::C1}\\
    C_{2l}(r)=C_{2l0}+\frac{1}{2l+1}\int_R^r s^{1-l}\alpha_l(s)\mathrm{d}s,\label{app::C2}
\end{align}
where the integration constants $C_{1l0}$ and $C_{2l0}$ must be calculated from additional boundary conditions.

\section{Solution for $\Psi$}

We now turn to the vector potential $\Psi^{(2)}$, and show how its solution is obtained in a condensed way. 
We start with Eq.~(\ref{eq::DGL_Psi})
\begin{align*}
    \nabla^4\vec{\Psi}^{(2)}=-\frac{\rho^{(0)}}{\eta_f+\eta_p}\nabla\times\langle\vec{v}^{(1)}\nabla\cdot\vec{v}^{(1)}+\vec{v}^{(1)}\cdot\nabla\vec{v}^{(1)}\rangle-\frac{\lambda_M}{\eta_f+\eta_p}\nabla\times\langle\nabla\cdot\mathbf{T}\rangle,
\end{align*}
and follow a similar treatment of expanding the inhomogeneity in Legendre polynomials. 

Starting with the first term on the right-hand side, we expand the argument of the rotation. The radial component is expanded in Legendre and the tangential component in associated Legendre polynomials
\begin{align}
    \langle\vec{v}^{(1)}\nabla\cdot\vec{v}^{(1)}+\vec{v}^{(1)}\cdot\nabla\vec{v}^{(1)}\rangle_r=\sum_{l=0}^\infty\beta_l(r)P_l(\mu)\label{eq::app::exp_beta},\\
    \langle\vec{v}^{(1)}\nabla\cdot\vec{v}^{(1)}+\vec{v}^{(1)}\cdot\nabla\vec{v}^{(1)}\rangle_\theta=\sum_{l=1}^\infty\delta_l(r)P_l^1(\mu),\label{eq::app::exp_delta}
\end{align}
where the expansion in associated Legendre polynomials
\begin{align}
    \delta_l(r)&=\frac{2l+1}{2l(l+1)}\int_{-1}^1\langle\vec{v}^{(1)}\nabla\cdot\vec{v}^{(1)}+\vec{v}^{(1)}\cdot\nabla\vec{v}^{(1)}\rangle_\theta ~P_l^1(\mu)\mathrm{d}\mu,
\end{align}
has a different normalization constant.

The expansion of the divergence of
the stress tensor $\mathbf{T}$ is shown in Appendix~\ref{App::D} and results in
\begin{align}
    \langle\nabla\cdot\mathbf{T}\rangle=\vec{e}_r\sum_{l=0}^\infty D_l^{(r)}(r)P_l(\mu)+\vec{e}_\theta\sum_{l=1}^\infty D_l^{(\theta)}(r)P_l^1(\mu).\label{eq::app::exp_DelT}
\end{align}
By inserting the expansions Eqs.~(\ref{eq::reihe_psi2}), (\ref{eq::app::exp_beta}), (\ref{eq::app::exp_delta}), and (\ref{eq::app::exp_DelT}) in Eq.~(\ref{eq::DGL_Psi}) and calculating the rotations explicitly, we arrive at
\begin{align}
    \Psi^{(2),\prime\prime\prime\prime}_l+\frac{4}{r}\Psi^{(2),\prime\prime\prime}_l-\frac{2l(l+1)}{r^2}\Psi_l^{(2),\prime\prime}+\frac{l(l+1)(l^2+l-2)}{r^4}\Psi^{(2)}_l=G_l(r),\label{eq::app::2nd_multipol_darstellung}
\end{align}
with the inhomogeneity
\begin{align}
    G_l(r)=-\frac{1}{(\eta_f+\eta_p)r}\left[\rho^{(0)}\left(\delta_l(r)+r\delta_l'(r)-\beta_l(r)\right)+\lambda_M\left(D_l^{(\theta)}(r)+rD_l^{(\theta)'}(r)-D_l^{(r)}(r)\right)\right].\label{eq::app::Gl}
\end{align}
The solution for $\Psi^{(2)}_l$ is now available by using variation of constants
\begin{align}
    \Psi^{(2)}_l(r)=\frac{C_{3l}(r)}{r^{l-1}}+\frac{C_{4l}(r)}{r^{l+1}}+r^lC_{5l}(r)+r^{l+2}C_{6l}(r),\label{eq::app::Psi_sol}
\end{align}
where the coefficients can be calculated from the integrals
\begin{align}
    C_{3l}(r)&=C_{3l0}+\frac{1}{2(2l-1)(2l+1)}\int_R^{r}s^{l+2}G_l(s)\mathrm{d}s,\label{app::C3}\\
    C_{4l}(r)&=C_{4l0}-\frac{1}{2(2l+1)(2l+3)}\int_R^{r}s^{l+4}G_l(s)\mathrm{d}s,\label{app::C4}\\
    C_{5l}(r)&=C_{5l0}-\frac{1}{2(2l-1)(2l+1)}\int_R^{r}s^{3-l}G_l(s)\mathrm{d}s,\label{app::C5}\\
    C_{6l}(r)&=C_{6l0}+\frac{1}{2(2l+1)(2l+3)}\int_R^{r}s^{1-l}G_l(s)\mathrm{d}s,\label{app::C6}
\end{align}
and the integration constants must be determined using boundary conditions.

\section{Expansion of 
$\nabla\cdot\mathbf{T}$}
\label{App::D}

We first expand the tensor $\mathbf{T}$ in Legendre polynomials and apply the tensor divergence explicitly after this. To this end, we introduce the strain rate tensor $\gamma$ as
\begin{align}
    \gamma^{(1)}=\frac{1}{2}\left(\nabla\vec{v}^{(1)}+(\nabla\vec{v}^{(1)})^T\right),\label{eq::app::strainrate}
\end{align}
and rewrite the first-order polymer stress in Eq.~(\ref{eq::first_order_polymer_stress_prelim}) as
\begin{align}
    \tau^{(1)}=\frac{2\eta_p}{1-i\lambda_M\omega}\gamma^{(1)}+\frac{\xi_p-2\eta_p/3}{1-i\lambda_M\omega}(\nabla\cdot\vec{v}^{(1)})\mathbbm{1}.\label{eq::app::tau1}
\end{align}
Additionally, we rewrite Eq.~(\ref{eq::T}) as
\begin{align}
    \mathbf{T}=\vec{v}^{(1)}\cdot\nabla\tau^{(1)}-2\tau^{(1)} \gamma^{(1)}-\left[(\nabla\vec{v}^{(1)})^T,\tau^{(1)}\right],\label{eq::app::T_commutator}
\end{align}
where $[\cdot,\cdot]$ denotes the commutator. The main advantage of this notation is its implementation in \textit{Mathematica}. The calculations are almost fully performed in the complex space, with the real part only taken as a last step. Therefore, there exist multiple different ways to write the imaginary part, which can greatly inflate or reduce the numerical load. Note, that while 
$\tau^{(1)}$ and $\gamma^{(1)}$ commute following Eq.~(\ref{eq::app::tau1}), this does not hold for the velocity gradient alone, so the commutator in Eq.~(\ref{eq::app::T_commutator}) is nonvanishing. 
Additionally, while the full tensor is orthogonal as by design of the convected derivative, the constituents themselves are not. This holds in particular for the product $\tau^{(1)} \gamma^{(1)}$. Even if the tensors are largely proportional to each other, the time average requires that the phase information stored in the complex proportionality constant is treated consistently. Both of these points were apparently disregarded in Ref.~\cite{VS_Doinikov}.

The tensor gradient in the first term 
of Eq.~(\ref{eq::app::T_commutator})
requires further consideration. It can be calculated using Christoffel symbols as
\begin{align}
    [\nabla\tau^{(1)}]_{rrr}&=\frac{\partial \tau^{(1)}_{rr}}{\partial r},&
    [\nabla\tau^{(1)}]_{\theta rr}&=\frac{1}{r}\left(\frac{\partial\tau^{(1)}_{rr}}{\partial\theta}-\tau^{(1)}_{r\theta}-\tau^{(1)}_{\theta r}\right),\label{eq::app::Christoffel}\\\nonumber
    [\nabla\tau^{(1)}]_{r\theta\theta}&=\frac{\partial\tau^{(1)}_{\theta\theta}}{\partial r},&
    [\nabla\tau^{(1)}]_{\theta\theta\theta}&=\frac{1}{r}\left(\frac{\partial\tau^{(1)}_{\theta\theta}}{\partial\theta}+\tau^{(1)}_{r\theta}+\tau^{(1)}_{\theta r}\right),\\\nonumber
    [\nabla\tau^{(1)}]_{r\phi\phi}&=\frac{\partial\tau^{(1)}_{\phi\phi}}{\partial r},&
    [\nabla\tau^{(1)}]_{\theta\phi\phi}&=\frac{1}{r}\frac{\partial\tau^{(1)}_{\phi\phi}}{\partial \theta},\\\nonumber
    [\nabla\tau^{(1)}]_{rr\theta}&=\frac{\partial\tau^{(1)}_{r\theta}}{\partial r},&
    [\nabla\tau^{(1)}]_{\theta r\theta}&=\frac{1}{r}\frac{\partial\tau^{(1)}_{r\theta}}{\partial\theta}+\frac{1}{r}\left(\tau^{(1)}_{rr}-\tau^{(1)}_{\theta\theta}\right),\\\nonumber
    [\nabla\tau^{(1)}]_{r\theta r}&=\frac{\partial\tau^{(1)}_{\theta r}}{\partial r},&
    [\nabla\tau^{(1)}]_{\theta\theta r}&=\frac{1}{r}\frac{\partial\tau^{(1)}_{\theta r}}{\partial\theta}+\frac{1}{r}\left(\tau^{(1)}_{rr}-\tau^{(1)}_{\theta\theta}\right),
\end{align}
where the first index has to be contracted with the velocity vector, and the last two determine the tensor component. Special care has to be taken as we are working with normalized unit vectors, as opposed to the typical notation of differential geometry.

Inserting Eqs.~(\ref{eq::app::strainrate}), (\ref{eq::app::tau1}), and (\ref{eq::app::Christoffel}) in Eq.~(\ref{eq::app::T_commutator}), we can express $\mathbf{T}$ fully with the first-order velocities, and, therefore, expand it in Legendre polynomials
\begin{align}
    T_{rr}=\sum_{l=0}^\infty T^{(rr)}_l(r)P_l(\mu),\\
    T_{\theta\theta}=\sum_{l=0}^\infty T^{(\theta\theta)}_l(r)P_l(\mu),\\
    T_{\phi\phi}=\sum_{l=0}^\infty T^{(\phi\phi)}_l(r)P_l(\mu),\\
    T_{r\theta}=\sum_{l=1}^\infty T^{(r\theta)}_l(r)P_l^1(\mu),\\
    T_{\theta r}=\sum_{l=1}^\infty T^{(\theta r)}_l(r)P_l^1(\mu).
\end{align}
While these expansions can, as previously stated, be explicitly calculated, the calculation is performed with \textit{Mathematica}. With this result, the divergence can be calculated explicitly, yielding
\begin{align}
    \langle\nabla\cdot\mathbf{T}\rangle_r&=\sum_{l=0}^\infty D_l^{(r)}P_l(\mu),\\\label{d-koeff}
    \langle\nabla\cdot\mathbf{T}\rangle_\theta&=\sum_{l=1}^\infty D_l^{(\theta)}P_l^1(\mu),
\end{align}
where the coefficients $D_l^{(r)}$ and $D_l^{(\theta)}$ are given by
\begin{align}
    D_l^{(r)}(r)&=T_l^{(rr)'}(r)+\frac{2T_l^{(rr)}(r)-T_l^{(\theta\theta)}(r)-T_l^{(\phi\phi)}(r)-l(l+1)T_l^{(r \theta )}(r)}{r},\label{eq::app::Dlr}\\
    D_l^{(\theta)}(r)&=T_l^{(\theta r)'}(r)+\frac{2(l+1)T_l^{(\theta r)}(r)+(l+1)T_l^{(r \theta)}(r)+lT_l^{(\theta\theta)}(r)+T_l^{(\phi\phi)}(r)}{(l+1)r}-\frac{2l+1}{l(l+1)}\sum_{k=1}^{[l/2]}\frac{T_{l-2k}^{(\theta\theta)}(r)-T_{l-2k}^{(\phi\phi)}(r)}{r}.\label{eq::app::Dltheta}
\end{align}
Here, the second term of Eq. \ref{eq::app::Dltheta} results from the expansion
\begin{align}
    \frac{\mu P_l(\mu)}{\sqrt{1-\mu^2}}=\sum_{n=1}^\infty\left(-\frac{\delta_{nl}}{n+1}-\frac{2n+1}{n(n+1)}\sum_{k=1}^{[n/2]}\delta_{(n-2k)l}\right)P_n^1(\mu),
\end{align}
where $[\cdot]$ denotes the inside expression rounded down.

\section{Integration Constants \label{app::BCs}}
After performing the variation of constants for the potentials $\Phi^{(2)}$ and $\Psi^{(2)}$, there are still the integration constants $C_{il0}$ to determine, where the index $i$ stands for the 6 constants, both inside and outside; leading to 12 constants per order in total. Out of these, four can be determined using ambiguity, limiting behaviour, and boundary conditions each.

\subsection{Gauge of the Velocity Potentials}
When trying to determine the value of the constants $C_{il0}$, it is important to note that any conditions apply only to the velocity field, not the potentials themselves. In particular, 
\begin{align}
    \nabla\times\left[\hat{e}_\phi\, P_l^1(\mu)r^{-(l+1)}\right]=l\nabla\left[P_l(\mu)r^{-(l+1)}\right]\label{eq::app::C1C4}
\end{align}
shows that the contributions of the terms proportional to $C_{1l0}$ and $C_{4l0}$ have the same functional behavior, thus containing one internal degree of freedom, which has to be determined by the choice of on an appropriate gauge.

Similarly, from
\begin{align}
        \nabla\times\left[\hat{e}_\phi\, P_l^1(\mu)r^l\right]=-(l+1)\nabla\left[P_l(\mu)r^l\right],\label{eq::app::C2C5}
\end{align}
$C_{2l0}$ and $C_{5l0}$ are connected, leaving us with two choices of gauge both inside and outside of the sphere, resulting in four constants in total. In the following, we choose $C_{1l0,\mathrm{out}}=C_{2l0,\mathrm{in}}=0$ for two of these gauges, and specify the remaining two conditions when considering the limiting behavior.

Note that this works only in the case of $l\neq0$ in which both $\Phi^{(2)}$ and $\Psi^{(2)}$ are nonzero. However, as the Lagrangian streaming velocity has no contribution from the $l=0$ order, as shown in Appendix~\ref{App::no_0th}, 
we only calculate the potentials for $l\neq0$.

\subsection{Limiting Behaviour}\label{app::limitingbehaviour}
The streaming velocity field due to scattering should tend to zero for both $r\rightarrow 0$ and $r\rightarrow\infty$ to ensure physical behavior. At the center of the sphere, this results in the following conditions
\begin{align}
    C_{3l,\mathrm{in}}(0)&=0 \Rightarrow C_{3l0,\mathrm{in}}=-\frac{1}{2(2l-1)(2l+1)}\int_R^{0}s^{l+2}G_l(s)\mathrm{d}s\label{eq::app::C3l0},\\
    C_{4l,\mathrm{in}}(0)&=0\Rightarrow C_{4l0,\mathrm{in}}=\frac{1}{2(2l+1)(2l+3)}\int_R^{0}s^{l+4}G_l(s)\mathrm{d}s\label{eq::app::C4l0},\\
    C_{1l,\mathrm{in}}(0)&=0\Rightarrow C_{1l0,\mathrm{in}}=\frac{1}{2l+1}\int_R^0 s^{l+2}\alpha_l(s)\mathrm{d}s.\label{eq::app::C1l0}
\end{align}
However, there are some subtleties regarding these expressions. First, the condition for $C_{3l0,\mathrm{in}}$ is not mandatory for $l=1$, but seeing as the constant contribution to  $\Psi^{(2)}$ would have no effect on $\langle\vec{v}^{(2)}\rangle$, it is kept for consistency. Second, as discussed above, the contributions from $C_{1l0,\mathrm{in}}$ and $C_{4l0,\mathrm{in}}$ have the same functionality, so the condition is in fact only that a combination of $C_{1l0,\mathrm{in}}$ and $C_{4l0,\mathrm{in}}$ has to be chosen in such a way that the divergence at $r=0$ is canceled. 
We choose the gauge such that both integration constants prevent their respective divergence separately.

Additionally, the streaming generated by the scatterer should fall off to zero at infinity. Note that this explicitly only applies to the streaming generated by the scattered field; the bulk streaming does not follow this condition. By plugging Eqs~(\ref{eq::app::Phi_sol}) and (\ref{eq::app::Psi_sol}) into Eqs~(\ref{eq::V2r}) and (\ref{eq::V2theta}), we arrive at
\begin{align}
    lr^{l-1}C_{2l,\mathrm{sc}}(r)-l(l+1)r^{l-1}C_{5l,\mathrm{sc}}(r)-l(l+1)r^{l+1}C_{6l,\mathrm{sc}}(r)=0 \quad\mathrm{for}\quad r\rightarrow\infty,\\
    r^{l-1}C_{2l,\mathrm{sc}}(r)-(l+1)r^{l-1}C_{5l,\mathrm{sc}}(r)-(l+3)r^{l+1}C_{6l,\mathrm{sc}}(r)=0 \quad\mathrm{for}\quad r\rightarrow\infty,
\end{align}
where the subscript "$\mathrm{sc}$" indicates that only the contributions due to scattering are considered. From this we can conclude that $C_{6l,\mathrm{sc}}(r)=0$ for $r\rightarrow\infty$. The condition for $C_{2l,\mathrm{sc}}(r)$ and $C_{5l,\mathrm{sc}}(r)$ can not be separated, however, we choose the last remaining freedom of gauge such that both $C_{2l,\mathrm{sc}}(r)=0$ and $C_{5l,\mathrm{sc}}(r)=0$ for $r\rightarrow\infty$. This results in 
\begin{align}
    C_{2l0}&=-\frac{1}{2l+1}\int_R^\infty s^{1-l}(\alpha_l(s)-\alpha_{l,\mathrm{ac}}(s))\mathrm{d}s,\\
    C_{5l0}&=\frac{1}{2(2l-1)(2l+1)}\int_R^{\infty}s^{3-l}(G_l(s)-G_{l,\mathrm{ac}}(s))\mathrm{d}s,\\
    C_{6l0}&=-\frac{1}{2(2l+1)(2l+3)}\int_R^{\infty}s^{1-l}(G_l(s)-G_{l,\mathrm{ac}}(s))\mathrm{d}s,
\end{align}
where the index "$\mathrm{ac}$" indicates that only terms proportional to the incoming coefficients are included, that means the scattering coefficients are all set to zero, resulting in the pure bulk streaming terms.

When expanding the spherical Bessel functions explicitly, terms of the form $e^{kr}/r$ can appear in these integrals, which are only conditionally convergent. It is therefore preferable to calculate the far field integrals analytically. In the numerical implementation, analytical expressions are used from a threshold value of $R_\mathrm{thr}$ with $\Re(k_f)R_\mathrm{thr}\sim 1$ where they are stable, and numerical integration is employed for $R\leq r\leq R_\mathrm{thr}$.  

\subsection{Boundary Conditions}
The last four integration constants are calculated using boundary conditions on the surface of the sphere. As these boundary conditions are now to second order, this includes the first-order oscillation of the surface. Instead of considering the oscillation explicitly, the conditions are calculated on the zeroth-order surface, including a Taylor expansion of the movement, which, for the velocity, is equivalent to considering the Lagrangian velocity. The resulting conditions are 
\begin{align}
    V_{Lo,\theta}&=V_{Li,\theta},\\
    \left\langle t^{(2)}_{o}\right\rangle+\left\langle \frac{i}{\omega}\left[\vec{v}^{(1)}_{o}\cdot \nabla \sigma^{(1)}_{o}\right]_{r\theta}\right\rangle&=\left\langle t^{(2)}_{i}\right\rangle+\left\langle \frac{i}{\omega}\left[\vec{v}^{(1)}_{i}\cdot \nabla \sigma^{(1)}_{i}\right]_{r\theta}\right\rangle,\label{eq::app::2ndtangential_stress_BC}\\
    V_{Li,r}&=0,\label{eq::app::BC_V2rLi}\\
    V_{Lo,r}&=0\quad\mathrm{at}\,\,r=R,\label{eq::app::BC_V2rLo}
\end{align}
where $\vec{V}_L$ with index L is the Lagrangian streaming velocity, defined as the Eulerian velocity $\langle\vec{v}^{(2)}\rangle$ plus the Stokes drift, 
cf.~Eq.~(\ref{eq::app::stokesdrift}),
and $t^{(2)}$ denotes the second order tangential stress
\begin{align}
    t^{(2)}= (\eta_f+\eta_p)\left(\frac{1}{r}\frac{\partial v^{(2)}_{r}}{\partial\theta}-\frac{v^{(2)}_{\theta}}{r}+\frac{\partial v^{(2)}_{\theta}}{\partial r}\right)-\lambda_M\langle T\rangle_{r\theta}.  \label{eq::app::second_order_tangential_stress}
\end{align}
Note that the radial Lagrangian streaming velocity has to vanish on the surface, because the boundary is considered stationary in the second order problem. As the streaming velocity is treated as a steady flow, any nonzero radial velocity component would result in a net displacement of the boundary position.

\section{Vanishing Zeroth Order Lagrangian Velocity}\label{App::no_0th}
The solution for the Eulerian streaming velocity in Eq.~(\ref{eq::V2rReihe}) allows for a non-vanishing contribution for $l=0$. As this is a purely extensional steady flow, its interpretation is problematic. However, the particle transport is characterized by the Lagrangian velocity. Therefore, to ensure a physical solution, we require $V_{Lr0}(r)=0$. By inserting Eq.~(\ref{eq::V2r}) in Eq.~(\ref{eq::V2rReihe}), we arrive at
\begin{align}
    \langle v^{(2)}_{r0}\rangle=-\frac{C_{10}(r)}{r^{2}}+V_{Sr0}\overset{!}{=}0,
\end{align}
where $V_{Sr0}$ is the zeroth order of the radial component of the Stokes drift. As the boundary conditions Eqs.~(\ref{eq::app::BC_V2rLi}) and (\ref{eq::app::BC_V2rLo}) ensure the condition is fulfilled on the surface, we can instead investigate the derivative
\begin{align}
    \frac{\partial}{\partial r}\left[r^2 V_{Sr0}(r)\right]\overset{!}{=}-r^2\alpha_0(r).
\end{align}
The inhomogeneity is given by 
\begin{align}
    \alpha_0(r,\mu)=&\int_{-1}^1\frac{1}{2\omega}\Re\Bigg\{i\nabla\cdot\left((\nabla\cdot\Vec{v}^{(1)})\right)\Vec{v}^{(1)*}\Bigg\}\mathrm{d}\mu\\
    =&\frac{1}{4\omega}\Re\Bigg\{i\sum_{l=0}^{\infty}\left[v^{(1)\prime\prime}_{rl}+\frac{2v^{(1)\prime}_{rl}}{r}-\frac{2v^{(1)}_{rl}}{r^2}+\frac{l(l+1)}{r^2}v^{(1)}_{\theta l}-\frac{l(l+1)}{r}v^{(1)\prime}_{\theta l}\right]v^{(1)*}_{rl}\frac{2}{2l+1}\nonumber\\
    +& i\sum_{l=1}^{\infty}\left[\frac{v^{(1)\prime}_{rl}}{r}+\frac{2v^{(1)}_{rl}}{r^2}-\frac{l(l+1)}{r^2}v^{(1)}_{\theta l}\right]v^{(1)*}_{\theta l}\frac{2l(l+1)}{2l+1}\Bigg\}\label{eq::app::alpha0}
\end{align}
where opposed to Eq.~(\ref{eq::DGL_Phi}), we do not calculate the divergence of the first-order velocity explicitly in the first line, and insert the expansion of the first-order velocities Eqs.~(\ref{eq::app::vr1multipole}) and (\ref{eq::app::vtheta1multipole}) to arrive at the second line. Note that we can generalize the second sum to include the term $l=0$, as it is vanishing. \\
Additionally, from Eq.~(\ref{eq::app::stokesdrift})
\begin{align}
    V_{Sr}=\frac{1}{2\omega}\Re\left\{i\left[v^{(1)}_r\frac{\partial v_r^{(1)*}}{\partial r}+\frac{v^{(1)}_\theta}{r}\frac{\partial v_r^{(1)*}}{\partial\theta}-\frac{v^{(1)}_\theta v_\theta^{(1)*}}{r}\right]\right\},
\end{align}
which results in 
\begin{align}
        \frac{\partial}{\partial r}r^2V_{Sr0}=\frac{1}{4\omega}\Re\Bigg\{i\Bigg[&\sum_{l=0}^{\infty}\left(2rv^{(1)}_{rl}v^{(1)\prime *}_{rl}+r^2v^{(1)\prime}_{rl}v^{(1)\prime *}_{rl}+r^2v_{rl}v^{(1)\prime\prime *}_{rl}\right)\frac{2}{2l+1}\nonumber\\
        +&\sum_{l=0}^{\infty}\left(v^{(1)}_{\theta l}v^{(1)*}_{rl}+rv^{(1)\prime}_{\theta l}v^{(1)*}_{rl}+rv^{(1)}_{\theta l}v^{(1)\prime *}_{rl}\right)\frac{2l(l+1)}{2l+1}\Bigg]\Bigg\}.\label{eq::app::Vrs0}
\end{align}
Due to the real part enclosing both Eq.~(\ref{eq::app::alpha0}) and Eq.~(\ref{eq::app::Vrs0}), all summands in the expressions can be conjugated freely, and the two expressions are equal. Therefore, the Lagrangian streaming velocity has no contribution for $l=0$, and we drop this index from Eq.~(\ref{eq::V2rReihe}).
\end{widetext}

%\bibliography{sample}% Produces the bibliography via BibTeX.
%apsrev4-2.bst 2019-01-14 (MD) hand-edited version of apsrev4-1.bst
%Control: key (0)
%Control: author (8) initials jnrlst
%Control: editor formatted (1) identically to author
%Control: production of article title (0) allowed
%Control: page (0) single
%Control: year (1) truncated
%Control: production of eprint (0) enabled
\providecommand{\noopsort}[1]{}\providecommand{\singleletter}[1]{#1}%

\end{document}